\documentclass[preprint, aps, prl, superscriptaddress, nofootinbib] {revtex4-1} 
\usepackage[english]{babel}
\usepackage{amsfonts}
\usepackage{amsmath}
\usepackage[psamsfonts]{amssymb}
\usepackage{graphicx}
\usepackage{textcomp}
\usepackage{tabularx} 
\usepackage{color, colortbl}

\newcommand{\e}{\varepsilon}
\newcommand{\C}{{\mathcal{C}}}
\newcommand{\N}{{\mathcal{N}}}
\newcommand{\kBT}{k_{\rm B}T}
\newcommand{\DG}{\Delta G}
\newcommand{\kB}{k_{\rm B}}
\newcommand{\DGb}{\Delta G_{\rm bind}}
\newcommand{\DX}{\Delta G}
\newcommand{\DW}{\Delta W}

\definecolor{LightBlue}{rgb}{0.5,.5,1}
\definecolor{LightRed}{rgb}{1,.5,.5}
\definecolor{LightGreen}{rgb}{0.5,1,.5}

\begin{document}

\title{Experimental measurement of binding energy, selectivity and allostery using fluctuation theorems}
 
\author{J. Camunas-Soler}
\affiliation{Small Biosystems Lab, Departament de F\'{\i}sica de la
  Mat\`eria Condensada, Facultat de F\'isica, Universitat de Barcelona,
  Barcelona, Spain}
 \affiliation{CIBER de Bioingenier\'ia, Biomateriales
  y Nanomedicina, Instituto de Salud Carlos III, Madrid, Spain}
  \affiliation{Present address: Departments of Bioengineering and Applied Physics, 
  Stanford University, Stanford, California 94304, USA}
 \author{A. Alemany}
 \affiliation{Small Biosystems Lab, Departament de F\'{\i}sica de la
  Mat\`eria Condensada, Facultat de F\'isica, Universitat de Barcelona,
  Barcelona, Spain}
  \affiliation{CIBER de Bioingenier\'ia, Biomateriales
  y Nanomedicina, Instituto de Salud Carlos III, Madrid, Spain}
    \affiliation{Present address: Hubrecht Institute-KNAW and University Medical Center Utrecht, 3584 CT Utrecht, the Netherlands}
  \author{F. Ritort}
  \affiliation{Small Biosystems Lab, Departament de F\'{\i}sica de la
  Mat\`eria Condensada, Facultat de F\'isica, Universitat de Barcelona,
  Barcelona, Spain} \affiliation{CIBER de Bioingenier\'ia, Biomateriales
  y Nanomedicina, Instituto de Salud Carlos III, Madrid, Spain}

\begin{abstract}

Thermodynamic bulk measurements of binding reactions rely on the validity of the law of mass action and the assumption of a dilute solution. Yet important biological systems such as allosteric ligand-receptor binding, macromolecular crowding, or misfolded molecules may not follow these assumptions and require a particular reaction model. Here we introduce a fluctuation theorem for ligand binding and an experimental approach using single-molecule force-spectroscopy to determine binding energies, selectivity and allostery of nucleic acids and peptides in a model-independent fashion. A similar approach could be used for proteins. This work extends the use of fluctuation theorems beyond unimolecular folding reactions, bridging the thermodynamics of small systems and the basic laws of chemical equilibrium.

\end{abstract}

\maketitle

Binding energies are key quantities determining the fate of intermolecular reactions~\cite{stormo2010determining}. Bulk experimental approaches such as
surface plasmon resonance, isothermal titration
calorimetry and fluorescent ligand binding assays, allow
the extraction of binding energies ($\Delta G^0_{{\rm bind}}$) from
measurements of the dissociation constant ($K_d$) with accuracy
  $\sim$1 kcal/mol through the expression:
\begin{equation}
\Delta G^0_{{\rm bind}}=-k_{\rm B}T\log\left[K_d\right]~,
\label{eq:masact}
\end{equation} 
where $k_{\rm B}$ is the Boltzmann constant, and $T$ the temperature~\cite{leavitt2001direct,mcdonnell2001surface}. However, many ligands such as DNA-binding
  proteins display different binding modes with varying affinities, or
require the concerted action of several subunits, making quantitative measurements challenging~\cite{kalodimos2004structure,kim2013probing}.

Force techniques such as optical tweezers can be used to pull on
individual ligand-DNA complexes allowing detection of binding events
one-at-a-time (Figure~1{\bf a}, inset)~\cite{bustamante2003ten,junker2005influence,cao2007functional,koirala2011single,ainavarapu2005ligand,hann2007effect}. However, force-induced ligand unbinding usually
takes place in non-equilibrium conditions, and binding energies cannot be directly inferred from the
measured work values. The Crooks fluctuation theorem and the Jarzynski
equality~\cite{jarzynski1997nonequilibrium,crooks2000path} are tools to extract equilibrium
free energy differences from work distributions obtained far from
equilibrium, allowing the measurement of folding free
energies of nucleic acids and proteins, both from fully
equilibrated~\cite{liphardt2002equilibrium,collin2005verification,shank2010folding} and kinetic
states~\cite{maragakis2008differential,junier2009recovery,alemany2012experimental}. However, to date the use of fluctuation
theorems remains restricted to unimolecular reactions (e.g. folding).

Here we introduce a fluctuation theorem for ligand binding (FTLB) that allows us to directly extract binding energies of bimolecular or higher-order reactions from irreversible work measurements in
pulling experiments (see S1.1 in \cite{suppmaterials}). We first show how cyclic
protocols allow an unambiguous classification of experimental
pathways in relation to the initial and final state, which is an
essential step in the application of these theorems. We then apply the FTLB to directly
  verify the validity of the law of mass action for dilute ligand
  solutions. Next we use the FTLB to accurately measure specific and
nonspecific binding energies, as well as allosteric effects due to the
cooperative binding of ligand pairs. Finally, we show how the FTLB is also applicable to
extract binding energies to non-native structures (e.g. misfolded
states, prions, chaperones), a measurement inaccesible to most bulk
techniques~\cite{orte2008direct,fierz2012stability,yu2012direct}.

As a proof of principle we investigated the binding of the restriction
endonuclease EcoRI to a 30-bp DNA hairpin that contains its recognition
site (GAATTC) (see S1.2,S1.3 in \cite{suppmaterials}). Restriction endonucleases, which bind their
cognate sequences with high affinity, are a paradigm of protein-DNA
interactions~\cite{lesser1990energetic,jen1997protein}. In a
typical experiment, the hairpin is unfolded (refolded) by increasing
(decreasing) the distance ($\lambda$) between the optical trap and the
micropipette (Figure~1{\bf a}). In the absence of ligand,
the hairpin folds and unfolds in the force range $F_c\sim12-15$ pN. The binding of EcoRI
increases the stability of the hairpin leading to higher unfolding
forces ($\sim$23 pN). During a
pulling experiment, EcoRI binds DNA when the hairpin is folded. However, since there is no net change in molecular
extension upon binding/unbinding, the native ($N$) and bound ($B$)
states cannot be distinguished at low forces. In contrast, at forces above $F_c\sim12-15$ pN the
bound state ($B$) can be unambiguously distinguished from the unfolded
state ($U$), as the hairpin remains folded when the protein is bound but unfolds when it is unbound
(Figure~1{\bf a}, empty, blue and cyan dots respectively).

The FTLB is based on the extended fluctuation
relation~\cite{maragakis2008differential,junier2009recovery,alemany2012experimental}, and relates the work ($W$) performed along a
pulling protocol connecting the different EcoRI-hairpin binding states
($N$, $B$, $U$) to their thermodynamic free-energy differences. We performed cyclic protocols that start and end at a force $\sim$21 pN
(Figure~1{\bf b} inset and Fig. S1). This force is
well above $F_c\sim12-15$ pN, resulting in paths that connect
states $U$ and $B$, that are then classified into four different sets
according to their initial and final states ($U\rightarrow U$,
$U\rightarrow B$, $B\rightarrow U$, $B\rightarrow B$,
Figure~1{\bf c}). We repeatedly pulled the hairpin
and measured the partial work distributions and fraction of paths connecting states $U$
and $B$ ($P^{B\rightarrow U}(W)$, $P^{U\rightarrow B}(W)$,
$\phi^{B\rightarrow U}$, $\phi^{U\rightarrow B}$ respectively)
and extracted the free-energy difference between states $B$ and $U$, $\Delta
G_{BU}$ ($=G_U-G_B$) using the FTLB (see S1.4 in \cite{suppmaterials}):
\begin{equation}
\frac{\phi^{B\rightarrow U}}{\phi^{U \rightarrow
    B}}\frac{P^{B\rightarrow U}(W)}{P^{U\rightarrow
    B}(-W)}=\exp\left[\frac{W-\Delta G_{BU}}{k_{\rm
      B}T}\right]~~~~.\label{eq:FT}
\end{equation}

We performed experiments at different EcoRI concentrations, and
determined $\Delta G_{BU}$ from the work value ($\widetilde{W}$) at
which the partial work distributions cross ($P^{B\rightarrow
  U}(\widetilde{W})=P^{U\rightarrow B}(-\widetilde{W})$) by taking
$\Delta G_{BU}=\widetilde{W}+k_{\rm B}T\log\left(\phi^{U\rightarrow
  B}/\phi^{B \rightarrow U}\right)$ (Figure~1{\bf d}, S1.5 in \cite{suppmaterials}). The term $\Delta G_{BU}$ includes all the energetic contributions
involved in going from $B$ to $U$ (e.g. binding energy, conformational changes,
elastic terms, see S1.6 in \cite{suppmaterials}). By subtracting the elastic contributions and the energy
of formation of the hairpin from the measured $\Delta G_{BU}$ value, we
extract the binding energy at zero force ($\Delta G_{{\rm bind}}$) at
different EcoRI concentrations (Figure~1{\bf e} and
Tables S1, S2). As shown in Figure~1{\bf e},
$\Delta G_{{\rm bind}}$ follows the law of mass action
(Eq. \ref{eq:masact}), $\Delta G_{{\rm
    bind}}=\Delta G_{{\rm bind}}^0+k_{\rm B}T\log(C/C_0)$ with $\Delta
G^0_{{\rm bind}}=26\pm0.5$ $k_{\rm B}T$, providing a direct test of its
validity. This value is independent on the start/end force of the cyclic protocol and relies on a correct classification of paths (Fig. S2-S3). We also performed titration experiments with varying NaCl concentration showing
that EcoRI binding energy has a pronounced salt-dependency with slope
$m_{\rm [NaCl]}=-11\pm 2$ $k_{\rm B}T$ (Figure~1{\bf
  f} and Tables S3, S4), in agreement with previous bulk experiments ~\cite{terry1983thermodynamic,koch2002probing}. Finally, we repeated experiments with hairpins
containing non-cognate DNA sequences which did not show binding in the
same range of EcoRI concentrations, proving the specificity of the
interaction~\cite{bustamante2003ten,koirala2011single}.

To further test the validity of Eq.~\ref{eq:FT}, we investigated a model
system consisting of a short oligonucleotide of 10 bases that binds a
DNA hairpin. The
oligonucleotide can bind the substrate by base-pairing complementarity
when the hairpin is in the unfolded ($U$) state, thereby inhibiting the
refolding of the hairpin at low forces. At forces below the critical
force range of the hairpin ($F_c\sim8-10$ pN), the oligo-bound state
($B$) competes with the formation of the native hairpin ($N$), and
states $B$ and $N$ can be distinguished due to their different molecular
extension (Figure~2{\bf a}). To apply
Eq.~\ref{eq:FT}, we considered cyclic protocols that start and end at a
force lower than the range $F_c$ (Figure~2{\bf b}). From the measured partial work
distributions and fractions of paths connecting $N$ and $B$
(Figure~2{\bf c}) we extracted the binding energies at zero force
($\Delta G_{{\rm bind}}$) (Figure~2{\bf d} and
Tables S5, S6). Measured binding energies again follow the law of mass action
with $\Delta G^0_{{\rm bind}}=22\pm1$ $k_{\rm B}T$
(Figure~2{\bf d}). This agrees
with theoretical predictions using the nearest-neighbour
model ($\Delta G^0_{\rm th}=22$ $k_{\rm
  B}T$)~\cite{santalucia1998unified,huguet2010single} and
equilibrium experiments performed at the coexistence force of the
hairpin, where hopping due to binding/unbinding is
observed (Figs. S4-S6 and Table S7). The
inclusion of the ratio $\phi^{N\rightarrow B}/\phi^{B \rightarrow N}$
(Figure~2{\bf d}, inset) is essential to recover the
correct binding energies. Previous attempts to derive binding energies using unidirectional
work measurements and the Jarzynski equality did not
account for concentration-dependent effects in the chemical potential
that are essential in Equation~\ref{eq:FT}~\cite{koirala2011single} .

To prove the general power of the method, we studied
echinomycin, a small DNA bis-intercalator with selectivity for CG
steps~\cite{van1984echinomycin} that binds contiguous ACGT
sites cooperatively~\cite{bailly1996cooperativity}. We performed
experiments with a 12-bp DNA hairpin containing a single CG-step (SP
hairpin) that shows rupture forces in
the range $F_c\sim6-8$ pN (Figure~3{\bf a} and Fig. S7). In the presence of echinomycin the histogram of rupture forces is shifted
to higher values and shows a bimodal distribution, indicating two binding modes:
a high-affinity binding to the specific CG-site (high-force peak,
$\sim18$ pN), and a low-affinity binding to other non-specific sites
(low-force peak, $\sim12$ pN). To confirm this, we pulled a hairpin in
which we removed the specific binding site by inverting the CG-motif
(NSP hairpin). In the presence of
ligand only the low affinity peak is observed
(Figure~3{\bf a}).

To extract the binding energy of each mode, we performed cyclic protocols
that start at a force high enough to discriminate both binding
modes: we used $\sim18$ pN ($\sim13$ pN) for the SP (NSP) hairpin in order
to extract both the specific and nonspecific binding energy of the ligand. In
this way, we obtained paths connecting states $B$ and $U$, and extracted
the binding energy of the specific and nonspecific
modes (Tables S8-S11).  For both binding modes, $\Delta G_{{\rm
    bind}}$ follows the law of mass action with $\Delta G^0_{{\rm bind,
    SP}}=20.0\pm0.8$ $k_{\rm B}T$ and $\Delta G^0_{{\rm bind,
    NSP}}=13.2\pm0.5$ $k_{\rm B}T$ (Figure~3{\bf b}), which
give affinities of 2 nM and 1.8 $\mu$M respectively
(Eq. \ref{eq:masact}). This measurement of an affinity in the nM range
for the specific binding is compatible with quasi-equilibrium experiments (Fig. S8) and improves
previous studies where accurate measurements could not be obtained due
to the concurrent action of both modes~\cite{leng2003energetics}.

The FTLB allows us to go beyond
free-energy measurements of single ligands, and measure allosteric
effects between ligands binding at nearby positions~\cite{kim2013probing} .  For this, we
designed hairpin $NC$ which contains two ACGT sites separated by 2 bp
(Figure~3{\bf c}). The simultaneous binding of two ligands
can be distinguished from the binding of a single ligand from the force
rips observed in the force-distance curve (Fig. S9).  By applying the
FTLB we extracted the binding energy per ligand in the single
  and double bound states, and found that binding is
favoured by the presence of a neighbouring ligand. The FTLB allows us to quantitatively test
the distance-dependence of this allosteric effect by performing a differential measurement of
binding energies with hairpin $C$, which contains two contiguous sites (Figure~3{\bf c}, Tables S12-S14). The binding energy per ligand we obtain in the double bound state in hairpin $C$ is
$2.4\pm0.5$ $k_{\rm B}T$ higher than in hairpin $NC$, providing a direct
experimental measurement of cooperativity effects in ligand pairs as a
function of their distance.

Single-molecule manipulation is particularly suited to observe the formation of misfolded structures
(e.g. prions, amyloids)~\cite{yu2012direct,heidarsson2014direct}, but
methods to characterize binding to these species are currently lacking. By applying the FTLB
it is possible to extract the binding energy to these kinetically
stabilized non-native structures. By using a DNA hairpin
with two binding sites separated by 4bp, we observe the formation of a
misfolded structure consisting of two short (4bp) hairpins in series
(Figure 3{\bf d}, hairpin M). Such an off-pathway kinetic
state is unobservable in the absence of ligand due to its low energy of
formation, however it is kinetically stabilized by the binding of the
ligand. We applied Eq.~\ref{eq:FT} by choosing a starting point of the
cyclic protocol where the native-bound and misfolded-bound conformations
are distinguishable ($F\sim10$ pN), and found that the energy of binding
to both configurations are equal ($\Delta G_{\rm bind,M-N} = 2\pm1$
$k_{\rm B}T$, Fig. S10 and Tab. S15, S16). 

In this work, we have introduced a fluctuation theorem for ligand
binding (FTLB) to directly determine binding energies as a
function of ligand concentration in single-molecule experiments. Using
different biomolecular systems of increasing complexity we provide a
single-molecule verification of the law of mass action, and show
how the FTLB can account for mass exchange between a molecular
system and the environment. We can resolve binding energies to specific and
non-specific sites with affinities spanning six orders of
magnitude. The FTLB provides a direct experimental
measurement of binding energies without assuming any model or reaction
scheme, which is particularly useful in cases where the law of mass
action does not hold. To show this, we applied the FTLB in two
situations where this may happen: the cooperative binding of multiple
ligands to the same substrate and the stabilization of kinetic
structures through ligand binding - both measurements inaccessible to
bulk methods and relevant to many interactions between proteins and ligands.

The use of an inherently non-equilibrium method to obtain equilibrium
binding energies also grants access to molecular interactions that
equilibrate over very long timescales (e.g. nucleosome assembly) and
that can only be currently measured by indirect techniques such as
competition assays~\cite{leavitt2001direct,fierz2012stability,thaastrom2004histone}. The FTLB relates work measurements to binding energies without making any assumption on reaction kinetics or the ideal solution limit. Therefore it might be also used to test the explicit breakdown of the law of mass action in conditions where it is not applicable, for instance in crowded environments, where ligands exhibit compartmentalized dynamics due to steric hindrance interactions~\cite{schnell2004reaction}. Lastly, the applicabilty of the FTLB
is not restricted to biomolecular reactions, and might be directly
applied to other interacting systems that can only be explored through
non-equilibrium methods.

\section{Acknowledgements}
\label{sec:Acknow}
All authors acknowledge funding from grants ERC MagReps 267 862, FP7
grant Infernos 308850, Icrea Academia 2013 and FIS2013-47796- P.

All data used in this study are included in the main text and in the supplementary materials.

\section{Author Contributions}
J.C.-S. and A.A. equally contributed to this work.
\label{sec:Author}

\newpage
\section{Figure Captions}
\subsection{Figure 1}

\includegraphics[scale=1]{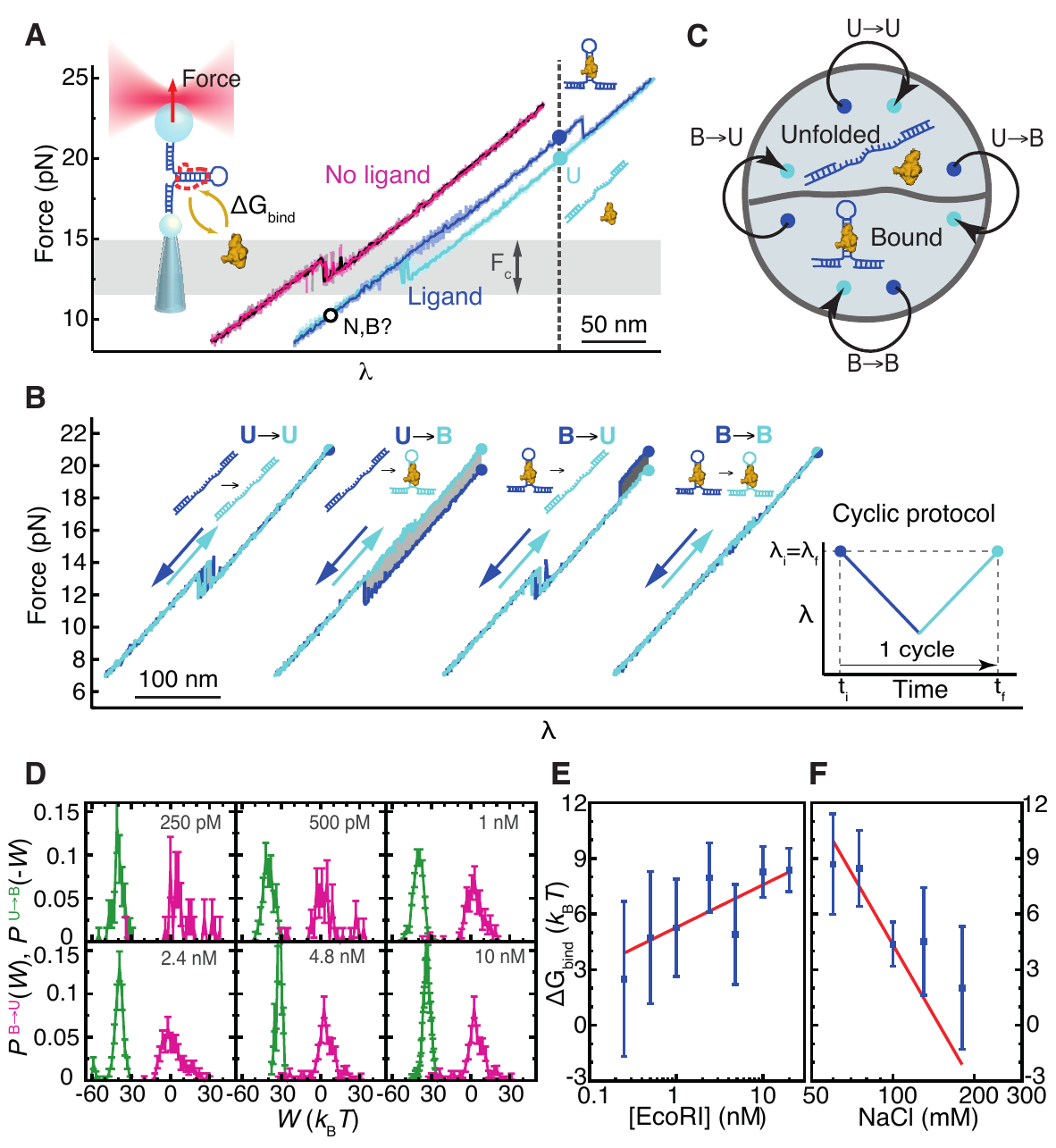}

{\bf EcoRI binding to DNA.} {\bf (a)} Unfolding/refolding force-distance
curves of a DNA hairpin in the absence (magenta/black) and presence
(blue/cyan) of EcoRI protein. The bound ($B$) and unfolded ($U$) states
are discriminated at high force by the presence of two distinct
force branches. {\bf (b)} Cyclic pulling
curves classified according to their initial (blue dot) and final state
(cyan dot) that start and end at a high force ($\sim$21 pN). Work equals
the enclosed area between the two curves and is shown in dark/light gray
for positive/negative values. {\bf (c)} Paths of a non-equilibrium cyclic protocol
connecting different initial and final states. {\bf (d)} Partial work distributions of
$U\rightarrow B$ (green) and $B\rightarrow U$ (magenta) transitions at
different EcoRI concentrations. {\bf (e)} Binding energy of EcoRI (blue) and fit to the law of mass action
(red line) at (130 mM $\rm{Na}^{+}$, $25^{\circ}$C, $C_0=1$ M). {\bf (f)} Binding energy of EcoRI at
varying [NaCl] (1 nM EcoRI). Error bars were obtained from
bootstrap using 1000 re-samplings of size N (N is total number of pulls for each condition shown in Tables S1 and S3). \newpage

\subsection{Figure 2}

\includegraphics[scale=1]{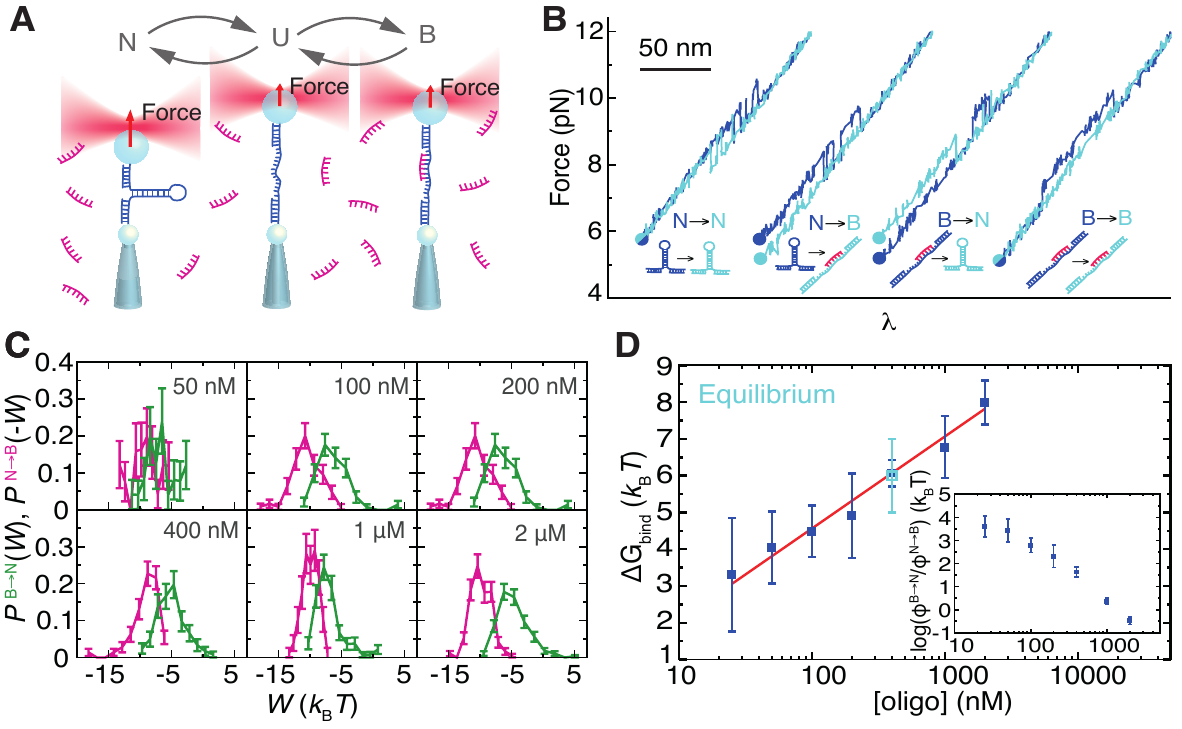}

{\bf Oligo binding to DNA.} {\bf (a)} Scheme of native ($N$),
unfolded ($U$) and oligo bound ($B$) states. {\bf (b)} Cyclic pulling
curves that start and end at low forces ($\sim$6 pN) classified
according to their initial (blue dot) and final state (cyan dot). {\bf
  (c)} Partial work distributions of $B\rightarrow N$ (green) and
$N\rightarrow B$ (magenta) transitions. {\bf
  (d)} Binding energy of the 10-base oligo (blue) and fit to the law of mass
action (red line). The value obtained from hopping equilibrium
experiments at [oligo]=400 nM (see S1.7 in \cite{suppmaterials}) is shown in cyan. (Inset)
Contribution of the ratio $\phi^{N\to B}/\phi^{B\to N}$ to the binding
energy. Error bars were obtained from bootstrap as described in Fig. 1.
\newpage

\subsection{Figure 3}

\includegraphics[scale=1]{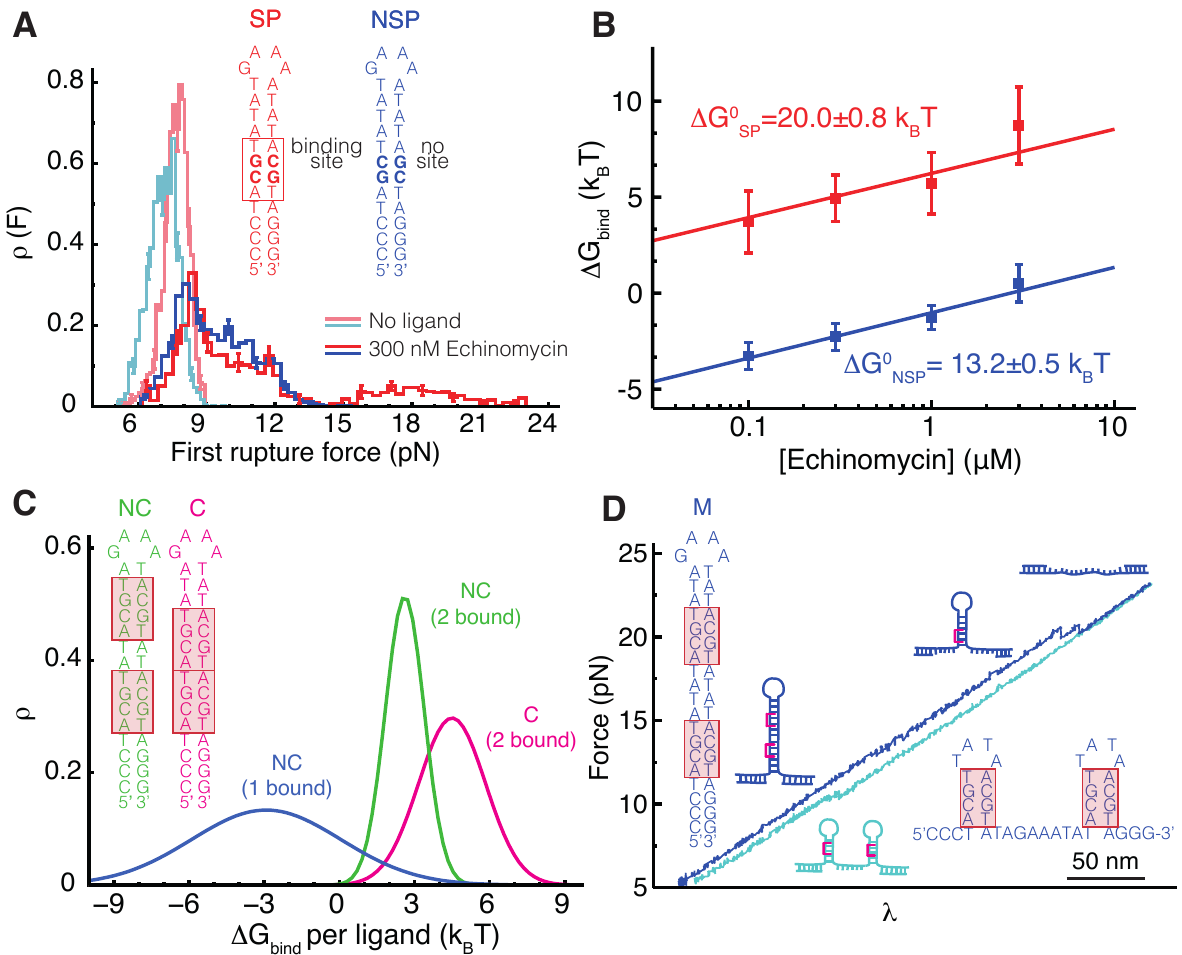}

{\bf Binding specificity, allostery and kinetic stabilization of misfolded
  states for the peptide Echinomycin.} {\bf (a)} First rupture force distribution
of hairpins SP (red) and NSP (blue) in the absence (light) and presence
(dark) of Echinomycin. {\bf (b)} Binding energy of Echinomycin to a
specific (red) and nonspecific (blue) site, and fit to the law of mass
action. {\bf (c)} Hairpins C and NC contain two specific binding sites
(red boxes) placed contiguously or
separated by 2-bp respectively. Binding energy per ligand when one
(blue) or two ligands (green) are bound to hairpins NC or C (magenta) ([Echninomycin]=3 $\mu$M). Gaussian distributions are reconstructed from mean and variance of measurements. The wider
distribution for the single bound state in hairpin NC (blue) is due to
the lower number of paths reaching this state at high ligand
concentration, increasing measurement error. {\bf (d)} Pulling cycle of hairpin M in the presence of Echinomycin. The unfolding curve (blue)
shows two force rips at $F\sim20$ pN corresponding to the unbinding of
two ligands bound to specific sites. In the refolding curve (cyan), the
hairpin does not fold back to the native state, and misfolds into a
kinetically stabilized configuration of longer molecular extension
($\sim40\%$ of refolding curves at [Echninomycin]=10
$\mu$M). Error bars were obtained from bootstrap as described in Fig. 1.
\newpage

\section{Materials and Methods}

\subsection{Mathematical proof of the Fluctuation Theorem for Ligand Binding (FTLB)}

The FTLB is derived following the same steps as in \cite{junier2009recovery}. 
Consider a system with a fluctuating number of particles $N$, which correspond to the ligand molecules. The system evolves under an experimental protocol $\lambda(t)$, where $\lambda$ denotes the control parameter and in our case corresponds to the position of the optical trap relative to the pipette. We discretize in time the protocol as $\lambda(t)=\{\lambda_0, \lambda_1,\dots,\lambda_{t_f}\}$, 
where $\lambda_i$ ($i=0,1,\dots,t_f$) denotes the value of $\lambda$ at the time of the protocol $t=i\Delta t$ (being $\Delta t$ the time discretization unit), and $t_f$ denotes the duration of the protocol. 
Along the protocol $\lambda(t)$ the system follows a given trajectory $\Gamma$, where a sequence of configurations $\C$ are sampled. The trajectory can be discretized as $\Gamma = \{ \C_0, \C_1, \dots, \C_{t_f}\}$. 
Each configuration $\C_i$ ($i\in 0, 1, \dots, t_f$) is characterized by the number of particles, $N_i$, and the degrees of freedom of each particle.

The equilibrium probability to be in a given configuration $\C_i$ at $\lambda$ can be written, according to the grand-canonical ensemble, as:
\begin{align}
\label{eq: total equilibrium}
 P^{\rm eq}(\C_i)=\frac{z^{N_i}e^{-\beta E_\lambda(\C_i)}}{Z^{GC}},& &Z^{GC} = \sum_{N_i}\sum_{\mathcal{C}_i(N_i)}z^{N_i} e^{-\beta E_\lambda(\mathcal{C}_i)}
\end{align}
where $\beta=(\kBT)^{-1}$ (being $\kB$ the Boltzmann constant and $T$ the absolute temperature), $z$ is the fugacity of the system (equal to $e^{\beta\mu}$, being $\mu$ the chemical potential of the ligand molecules), $Z^{GC}$ is the grand canonical partition function, and $E_\lambda(\C_i)$ is the energy of the configuration $\C_i$ at $\lambda$.

We suppose that the dynamics of the system satisfy the following detailed balance condition: 
\begin{equation}
 \frac{P(\C_t\to \C_{t+1})}{P(\C_{t+1}\to \C_t)}=z^{N_{t+1}-N_t}e^{-\beta\left(E_{\lambda(t+1)}(\C_{t+1})-E_{\lambda(t+1)}(\C_t)\right)}.
\end{equation}
Therefore, the probability of the system to follow a given trajectory $\Gamma$ (without imposing any initial and final configuration), and the probability to follow its time reversed $\hat{\Gamma}$ is: 
\refstepcounter{equation}
\label{eq: 1}
\[
P(\Gamma)= \prod_{t=0}^{t_f-1}P(\C_t\to \C_{t+1}),
\qquad \text{ } \qquad 
\hat{P}(\hat\Gamma)= \prod_{t=0}^{t_f-1}P(\hat \C_t\to \hat \C_{t+1}), 
\tag{\theequation a,b}\label{eq: 2}
\]
where $\hat \C_t=\C_{t_f-t}$.

We assume that in the forward protocol $\lambda(t)$ the system starts in partial equilibrium at $\C_0$, while in the reversed protocol $\hat\lambda(t)$ it starts in partial equilibrium at $\hat \C_0=\C_{t_f}$ (and $\hat\lambda(0)=\lambda(t_f)$). 
The partial equilibrium probability density function of a given configuration $\C_i\in S$, where $S$ is a subset of configurations accessible by the system, can be written as \cite{junier2009recovery}: 

\begin{align}\label{eq: partial}
 P^{\rm eq}_{S}(\C_i)&=\chi_{S}(\C_i)\frac{z^{N_i}e^{-\beta E(\C_i)}}{\sum_{N_i}\sum_{\C_i\in S}z^{N_i}e^{-\beta E(\C_i)}}=\chi_{S}(\C_i)P^{\rm eq}(\C_i)\frac{Z^{GC}}{Z^{GC}_{S}},
\end{align}
where $\chi_{S}(\C_i)$ is equal to one if $\C_i\in S$ and zero otherwise, and $Z^{GC}_S$ is the grand canonical partition function restricted to the subset $S$.

Suppose that the system starts in non-equilibrium conditions. Particularly, the system starts in partial equilibrium  at the kinetic state $S_0$ in the forward trajectory and at at the kinetic state $S_{t_f}$ in the reversed one. A kinetic state is a partially equilibrated region of configurational space, meaning that during a finite amount of time the system is confined and thermalized within that region. This is mathematically described by a Boltzmann-Gibbs distribution restricted to configurations contained in that region. Then:
\begin{subequations}
\begin{align}
\frac{P^{\rm eq}_{\lambda(0),S_0}(\C_0)P(\Gamma)}{P^{\rm eq}_{\lambda(t_f),S_{t_f}}(\C_{t_f})\hat P(\hat\Gamma)} &=
\frac{\chi_{S_0}(\C_0)P^{\rm eq}_{\lambda(0)}(\C_0)Z^{GC}_{\lambda(0)}}{Z^{GC}_{\lambda(0),S_0}}
\frac{Z^{GC}_{\lambda(t_f),S_{t_f}}}{\chi_{S_{t_f}}(\C_{t_f})P^{\rm eq}_{\lambda(t_f)}(\C_{t_f})Z^{GC}_{\lambda(t_f)}}
\frac{P(\Gamma)}{\hat P(\hat\Gamma)}\\
&=\frac{\chi_{S_0}(\C_0)Z^{GC}_{\lambda(t_f),S_{t_f}}}{\chi_{S_{t_f}}(\C_{t_f})Z^{GC}_{\lambda(0),S_0}}e^{\beta W(\Gamma)},
\end{align}
\end{subequations}
where:
\begin{equation}
 \label{eq: work}
W(\Gamma)=\sum_{t=0}^{t_f-1}\left(E_{\lambda(t+1)}(\C_{t})-E_{\lambda(t)}(\C_t)\right),
\end{equation}
is as the work exerted upon the system along the forward process.

Now we compute the average of the quantity $\mathcal{O}e^{-\beta W}$ over the forward trajectories that start in partial equilibrium in a configuration $\C_0\in S_0$ and end in $\C_{t_f}\in S_{t_f}$. Therefore:
\begin{align}
 \langle \mathcal{O}e^{-\beta W} \rangle^{S_0\to S_{t_f}} &= \frac{\sum_\Gamma P^{\rm eq}_{\lambda(0), S_0}(\C_0)P(\Gamma)\chi_{S_{t_f}}(\C_{t_f}) \mathcal{O}(\Gamma)e^{-\beta W(\Gamma)}}{\sum_\Gamma P^{\rm eq}_{\lambda(0), S_0}(\C_0)P(\Gamma)\chi_{S_{t_f}}(\C_{t_f})}\\
&=\frac{\phi^{S_{t_f}\to S_0}}{\phi^{S_0\to S_{t_f}}} \frac{Z^{GC}_{\lambda(t_f),S_{t_f}}}{Z^{GC}_{\lambda(0),S_0}}
\frac{\sum_{\Gamma}P^{\rm eq}_{\hat\lambda(t_0),\hat S_{0}}(\hat \C_{0})\hat P(\hat\Gamma)\chi_{\hat S_{t_f}}(\hat \C_{t_f}) \hat{\mathcal{O}}(\hat\Gamma)}{\sum_{\Gamma}P^{\rm eq}_{\hat\lambda(t_0),\hat S_{0}}(\hat \C_{0})\hat P(\hat\Gamma)\chi_{\hat S_{t_f}}(\hat \C_{t_f})}\\
&=\frac{\phi^{S_{t_f}\to S_0}}{\phi^{S_0\to S_{t_f}}} \frac{Z^{GC}_{\lambda(t_f),S_{t_f}}}{Z^{GC}_{\lambda(0),S_0}} \langle \hat{\mathcal{O}}\rangle^{S_{t_f}\to S_0}.
\end{align}

By defining $\mathcal{O}(\Gamma)=\delta(W-W(\Gamma))$ we obtain the extended Crooks relation in the grand-canonical ensemble, or equivalently the Fluctuation Theorem for Ligand Binding (FTLB):
\begin{subequations}
\label{eq: crooks}
 \begin{align}
 \frac{\phi^{S_0\to S_{t_f}}}{\phi^{S_{t_f}\to S_0}}\frac{P^{S_0\to S_{t_f}}(W)}{P^{S_{t_f}\to S_0}(-W)}&=e^{\beta W}\frac{Z^{GC}_{\lambda(t_f),S_{t_f}}}{Z^{GC}_{\lambda(0),S_0}}\\
&=\exp\left[\beta \left(W-\Delta G_{S_0S_{t_f}}\right)\right],
\end{align}
\end{subequations}
where $\Delta G_{S_0S_{t_f}} =  G(\lambda(t_f),S_{t_f})- G(\lambda(0),S_0)$.

\subsection{Force spectroscopy experiments with optical tweezers}

Experiments are performed with a highly stable miniaturized dual-beam optical tweezers described in previous studies~\cite{huguet2010single}. The DNA hairpins are tethered between two beads by using short dsDNA handles (29-bp) that are differentially end-labelled with biotin and digoxigenin to attach each handle to a different bead (see Supp. Section~\ref{sec: Summary of DNA hairpins used in this work} for hairpin sequences and synthesis details)~\cite{forns2011improving}. Pulling speed is set at 190 nm/s in all the experiments. For the EcoRI experiments, longer dsDNA handles ($\sim$600-bp) are used to reduce nonspecific interactions mediated by the protein and beads. For these experiments, the microfluidics chamber was also coated with Poly(ethylene glycol) (PEG) to avoid protein loss due to nonspecific absorption on the glass surface~\cite{cheng2011single}. Experimental conditions for each interaction were chosen to be comparable to previous ensemble and single-molecule studies: for EcoRI (Hepes 10 mM pH7.5, EDTA 1 mM, NaCl 130 mM, BSA 0.1 mg/ml, 100 $\mu$M, DTT, 0.01\% NaN$_3$) and salt titrated in the range 60-180 mM NaCl; for the oligonucleotide (Tris 10 mM pH7.5, EDTA 1 mM, 100 mM NaCl, 0.01\% NaN$_3$); and for echinomycin (Tris 10 mM pH7.5, EDTA 1 mM, 100 mM NaCl, 2\% DMSO, 0.01\% NaN$_3$). All ligands were obtained from commercial sources and used without further purification: EcoRI (New England Biolabs, 100 U/$\mu$l, $\sim$800 nM dimer), oligonucleotide (Eurofins MWG Operon, HyPur grade), Echinomycin (Merck Millipore). Concentrations were confirmed using a spectrophotometric analysis for the oligonucleotide (extinction coefficient 97400 M$^{-1}$cm$^{-1}$ at $\lambda$=260 nm) and echinomycin (extinction coefficient 11500 M$^{-1}$cm$^{-1}$ at $\lambda$=325 nm), whereas for EcoRI we performed an electrophoretic mobility shift assay using previously described protocols~\cite{sidorova1996differences}. All experiments were performed at 25\textcelsius. In all the cases, the number of cycles obtained per molecule range between 20 and 1300. A minimum of 2 and a maximum of 10 molecules were pulled in each case (see tables \ref{tab: EcoRI N}, \ref{tab: ecori N sal}, \ref{tab: oligo N}, \ref{tab: echi N}, \ref{tab: echins N}, \ref{tab: echi c nc N} \ref{tab: echi m N}).

\newpage
\subsection{DNA hairpins used in this work}
\label{sec: Summary of DNA hairpins used in this work}

DNA hairpin sequences, secondary structure, binding sites, thermodynamic information and synthesis details:

\begin{minipage}{7.8cm}
 \includegraphics[width=7.7cm]{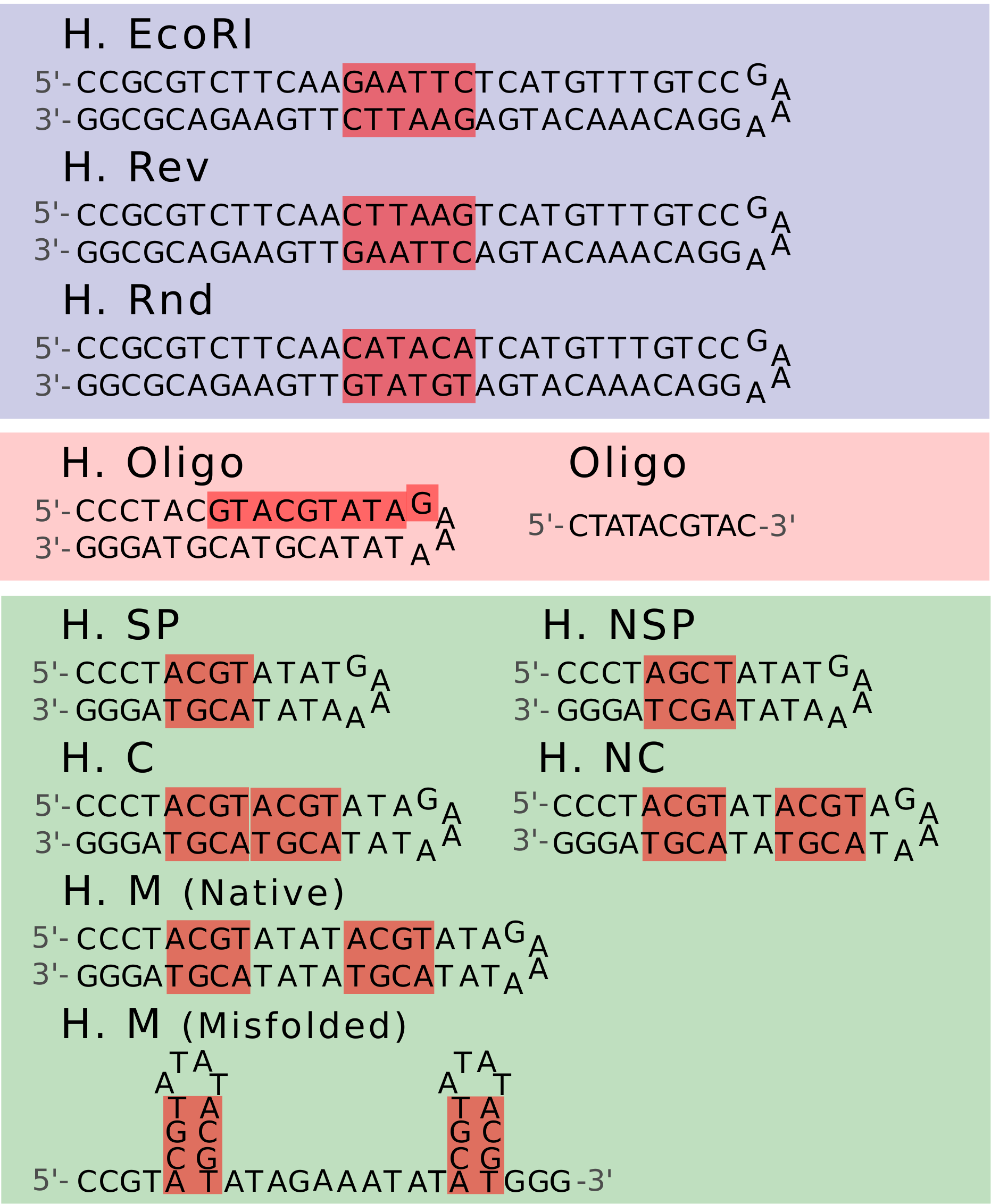}
\end{minipage}
\begin{minipage}{5cm}
 \begin{tabular}{lcc}
   Haipin & $\DG_{NU}^0$ ($\kBT$) & Handle length \\
  \hline\hline
  \rowcolor{LightBlue}
  EcoRI & 65 $\pm$ 3 & long \\
  \rowcolor{LightBlue}
  Rev   & 67 $\pm$ 2 & long \\
  \rowcolor{LightBlue}
  Rnd   & 68 $\pm$ 2 & long \\
  \hline\hline
  \rowcolor{LightRed}
  Oligo & 28 $\pm$ 1 & short \\
  \hline\hline
  \rowcolor{LightGreen}
  SP    & 19.7 $\pm$ 0.5 & short \\
  \rowcolor{LightGreen}
  NSP   & 19.4 $\pm$ 0.3 & short \\ 
  \rowcolor{LightGreen}
  C     & 18  $\pm$ 1   & short \\
  \rowcolor{LightGreen}
  NC    & 18   $\pm$ 1   & short \\
  \rowcolor{LightGreen}
  M ({\it Native}) & 33 $\pm$ 1   & short \\
  \rowcolor{LightGreen}
  \phantom{M} ({\it Misfolded}) & 10   $\pm$ 1   & \\
 \hline
 \end{tabular}
\end{minipage}

\noindent {\textbf{Blue-Top.}}~Hairpins used to study the binding of EcoRI to its recognition sequence (5'-GAATTC-3'). Hairpin EcoRI contains the specific binding site, indicated in red; Hairpin Rev contains its reversed sequence (red); and Hairpin Rnd contains a random sequence (red). We did not observe binding of EcoRI to hairpins Rnd and Rev in pulling experiments using these hairpins.\\
{\textbf{Red-Middle.}}~Hairpin (left) used to study the specific binding of an oligonucleotide (right) to its complementary ssDNA sequence. The binding site of the oligo is indicated in red.\\
{\textbf{Green-Bottom.}}~Hairpins used to study the binding of echinomycin to different binding motifs. Hairpin SP contains a specific binding site 5'-ACGT-3' (red). In hairpin NSP the specific binding site is removed by doing a sequence permutation (red). Hairpin C contains two contiguous binding sites (red), while hairpin NC contains two binding sites separated by two basepairs (red). Hairpin M contains two binding sites that are separated by four base pairs (hairpin M, native). In the presence of echinomycin, a misfolded structure containing two serially connected hairpins (hairpin M, misfolded) becomes kinetically stabilized by the binding of the ligand to two 4-bp hairpins containing the 5'-ACGT-3' motif (red).

\vspace{1em}

Mean values for the free energy of formation at zero force at 25\textcelsius\, and 130~mM NaCl (which are the experimental conditions unless stated otherwise) have been obtained by averaging over the results provided by the nearest-neighbor model and the unified oligonucleotide set of basepair free energies measured in bulk \cite{mfold,santalucia1998unified} and unzipping \cite{huguet2010single} experiments. Error bars are standard errors obtained between the two different estimations. To pull on the hairpins, handles of two different length are used in the experiments: short (29 bp) and long (500 bp) dsDNA handles. 

For the experiments with echinomycin and the oligonucleotide we used a short handle construct (total handle length: 58-bp). This short handles construct is better suited to study small ligands that might non-specifically bind to the dsDNA handles. Due to the short length of this construct, it can be synthesized by direct annealing and ligation of three partially complementary oligonucleotides that create the hairpin structure and dsDNA handles, as described in previous studies~\cite{forns2011improving}.

For the experiments performed with EcoRI we used a long handle construct to maintain a larger separation between the two beads at low forces (total handle length: 1322-bp). The synthesis is similar to the protocol described in~\cite{forns2011improving}. Briefly, the two handles are performed by PCR amplification of plasmid pBR322 to obtain DNA fragments that contain a restriction site for TspRI or Tsp45I respectively, but do not contain potential binding sites for the ligand (i.e. EcoRI). A biotin tag is introduced in one of the handles using a 5'-biotinylated primer on the PCR reaction. The other handle is tailed with digoxigenin-dUTP using the 3'-5' exonuclease activity of T4 DNA polymerase. After digestion of the labelled products with each enzyme, the TspRI/Tsp45I cohesive ends are used to anneal and ligate the handles to the the hairpin structure, that is assembled using oligonucleotides.

\subsection{Application of the FTLB: how to}
\label{sec: Application of the Fluctuation relation to our experiments}

To apply the FTLB to cyclic pulling protocols we follow the next steps:
\begin{enumerate}
 \item Identification of initial and final states in the non-equilibrium protocol. 
 
 In the particular cases studied in our work, initial and final states in the cyclic pulling experiments are:
\begin{itemize}
 \item Specific binding of EcoRI to dsDNA: $B$ and $U$ (high forces).
 \item Specific binding of an oligo to ssDNA: $B$ and $N$ (low forces).
 \item Specific binding of echinomycin to dsDNA: $B$ and $U$ (high forces).
 \item Non-specific binding of echinomycin to dsDNA: $B$ and $U$ (high forces).
 \item Cooperative binding of echinomycin to dsDNA: $B_2$, $B$ and $U$ (high forces).
 \item Kinetically-stabilized non-native structures due to the simultaneous binding of two echinomycin ligands to dsDNA, misfolded: $N$, $M$.
\end{itemize}

 \item Classification of trajectories.
 
 Each trajectory can be classified as a function of the initial and final state. For instance, in EcoRI experiments we have four types of trajectories: ($i$) start at $B$ and end at $U$; ($ii$) start at $B$ and end at $B$ ; ($iii$) start at $U$ and end at $B$ ; and ($iv$) start at $U$ and end at $U$.
 
 \item Obtain partial work distributions.
 
 In each cyclic trajectory, the work is calculated as the area enclosed in the trajectory. The partial work distribution is the histogram of work values restricted to each type of trajectory. 
 
 \item Obtain the grand-canonical partial partition function for the intial and the final state. 
 
 The partial partition function for a state $S$ is computed in the restricted subset of configurations $\C_j$ that characterize state $S$:
\begin{equation}
\label{eq: partial ZGC}
 Z^{GC}_S = \sum_{N_i\in S}\sum_{\C_j(N_i)\in S} z^{N_i}e^{-\beta E(\C_j)}, 
\end{equation}
where $N_i$ and $E(C_i)$ are the number of bound particles and the energy in configuration $C_i$, $z=e^{\beta\mu}$ is the fugacity, and $\mu$ is the chemical potential of the binding agent.

In the particular cases studied in this work these are:
\begin{itemize}
 \item Specific binding of EcoRI and echinomycin to dsDNA.
\refstepcounter{equation}
\label{eq: ZGC EcoRI}
\[
Z^{GC}_U = e^{-\beta G_U(\lambda_0)},
\qquad \text{ } \qquad 
Z^{GC}_B =  ze^{-\beta\left(\e+ G_{n}(\lambda_0)\right)}=e^{-\beta\left(\e-\mu+ G_{n}(\lambda_0)\right)}.
\tag{\theequation a,b}\label{eq: ZGC EcoRIab}
\]
The term $G_U(\lambda_0)$ is the free energy of the hairpin in state $U$ at $\lambda_0$. The term $G_{n}(\lambda_0)$ is the free energy of the hairpin at $\lambda_0$ with $n$ unfolded basepairs in the hairpin stem prior to the binding site, $\e$ is the binding energy of the ligand and $\mu$ is its chemical potential. We refer to $\mu-\e$ as the binding free energy $\DG_{\rm bind}$. 

\item Specific binding of a short oligo to its  complementary ssDNA sequence. 
\refstepcounter{equation}
\label{eq: ZGC oligo}
\[
Z^{GC}_N = e^{-\beta G_{N}(\lambda_0)},
\qquad \text{ } \qquad 
Z^{GC}_B =  ze^{-\beta\left(\e+G_U(\lambda_0)\right)}=e^{-\beta\left(\e-\mu+ G_U(\lambda_0)\right)},
\tag{\theequation a,b}\label{eq: 3}
\]
where $G_N(\lambda_0)$ and $G_U(\lambda_0)$ are the free energies of the hairpin at $\lambda_0$ in the folded ($N$) or unfolded ($U$) state respectively, and $\DG_{\rm bind}=\mu-\e$ is the binding free energy of the oligonucleotide. 

\item Non-specific binding of echinomycin to dsDNA.
\label{eq: ZGC EcoRI nons}
\[
Z^{GC}_U = e^{-\beta G_U(\lambda_0)},
\qquad \text{ } \qquad 
Z^{GC}_{B_i} =  ze^{-\beta\left(\e_i+ G_{n_i}(\lambda_0)\right)}=e^{-\beta\left(\e_i-\mu+ G_{n_i}(\lambda_0)\right)},
\tag{\theequation a,b}\label{eq: ZGC EcoRIab2}
\]
where $B_i$ ($i=1,\dots,\mathcal{N}$) is a possible binding state (echinomycin binding to any site in dsDNA),  $\e_i$ is the binding free energy at position $i$ and $G_{n_i}(\lambda_0)$ is the free energy of the hairpin  at $\lambda_0$ where the $n_i$ basepairs prior to the position binding site of echinomycin are unfolded.

The FTLB applies to each pair os states $B_i$ and $U$, therefore:
\begin{equation}
\label{eq: FT sum}
 \sum_{i=1}^\N\frac{\phi^{B_i\to U}}{\phi^{U\to B_i}}\frac{P^{B_i\to U}(W)}{P^{U\to B_i}(-W)}  = \sum_{i=1}^\N e^{\beta W}e^{-\beta\left(\mu-\e_i +\DG_{n_iU}(\lambda_0)\right)}.
\end{equation}
By assuming that $\phi^{B_i\to U} = \phi^{B\to U}$ and $\phi^{U\to B_i} = \frac{1}{\N}\phi^{U\to B}$, $\e_i=\e_n$ ($i=1\dots\mathcal{N}$) and:
\begin{align}
\label{eq: PBUW}
 P^{B\to U}(W) &= \frac{1}{\N}\sum_{i=1}^\N P^{B_i\to U}(W),
\end{align}
\begin{equation}
 P^{U\to B}(-W)\simeq P^{U\to B_i}(-W),~~\forall i,
\end{equation}
it can be shown that:
\begin{subequations}
\begin{align}
\label{eq: FT sum2}
 \frac{\phi^{B\to U}}{\phi^{U\to B}}\frac{P^{B\to U}(W)}{P^{U\to B}(-W)} &=  {e^{\beta W}}\frac {e^{-\beta(\mu-\e_n)}}{\N^2}\sum_{i=1}^\N e^{-\beta\DG_{n_iU}(\lambda_0)}\\
&=e^{\beta \left(W-\DG_{BU}\right)}
\end{align}
\end{subequations}
where $\DX_{BU} = \DGb-\kBT\log\sum_{i=1}^\N e^{-\beta\DG_{n_iU}(\lambda_0)} + 2\kBT\log \N$.

\item Simultaneous specific binding of two echinomycin molecules to sequential dsDNA sites (cooperativity).
\refstepcounter{equation}
\label{eq: ZGC Coop t}
\[
Z^{GC}_U = e^{-\beta G_U(\lambda_0)},
\qquad \text{ } \qquad 
Z^{GC}_{B^2} = z^2e^{-\beta(2\e +G_{n}(\lambda_0))},
\tag{\theequation a,b}\label{eq: ZGC Coop}
\]

\item Stabilization of misfolded states through simultaneous binding of two echinomycin molecules to sequential dsDNA sites.
\refstepcounter{equation}
\label{eq: ZGC Misfolded}
\[
Z^{GC}_B = z^2e^{-\beta\left(2\e+G_N(\lambda_0)\right)},
\qquad \text{ } \qquad 
Z^{GC}_M =  z'^2e^{-\beta\left(2\e'+G_M(\lambda_0)\right)}.
\tag{\theequation a,b}\label{eq: ZGC misfab}
\]

\end{itemize}

\item Plug everything into the FTLB. 

According to Eq.~\eqref{eq: crooks}, we write the FTLB for trajectories $A\to B$ ($B\to A$) as:
 \begin{equation}
 \label{eq: FR EcoRI}
\frac{\phi^{A\to B}}{\phi^{B\to A}}\frac{P^{A\to B}(W)}{P^{B\to A}(-W)} = e^{\beta W}\frac{Z^{GC}_B}{Z^{GC}_A}=e^{\beta (W-\DX_{AB})},
\end{equation}
Use the Bennett acceptance ratio method for a better estimation of $\DG_{AB}$ (section \ref{sec: Bennett}). 

\item Extract the elastic contributions to $\DG_{AB}$ in order to get the binding free energy $\DGb$ (Section \ref{sec: energy contributions}). 
\end{enumerate}

\subsection{Bennett acceptance ratio method}\label{sec: Bennett}

The Bennett acceptance ratio method is used to estimate the free-energy difference $\Delta G_{AB}$ between two states that satisfies Eq.~\eqref{eq: crooks} from non-equilibrium work measurements.
Given a set of $n_F$ ($n_R$) forward (reversed) work measurements $W_i$, it is shown in \cite{bennett,shirts} that the solution $u$ of the following transcendental equation:
\begin{equation}
 \frac{u}{\kBT} = -\log\left(\frac{\phi^{A\to B}}{\phi^{B\to A}}\right) + z_R(u) - z_F(u),
\end{equation}
where:
\begin{subequations}
\begin{equation}
z_R(u)=\log\frac{1}{n_R}\sum_{i=1}^{n_R}\left( \frac{e^{-\beta W_i}}{1+\frac{n_F}{n_R}e^{-\beta(W_i+u)}}\right)
\end{equation} 
\begin{equation}
z_F(u)=\log\frac{1}{n_F}\sum_{i=1}^{n_F}\left( \frac{1}{1+\frac{n_F}{n_R}e^{\beta(W_i-u)}} \right)
\end{equation}
\end{subequations}
minimizes the statistical variance of the free energy estimation for $u=\Delta G_{AB}$.

\subsection{Energetic contributions to the binding free energy}
\label{sec: energy contributions}

The free energy difference $\Delta G_{AB}$ obtained by using the FTLB contains the binding free energy of the ligand to the given substrate $\DGb$ plus elastic and thermodynamic energetic contributions of the experimental setup $\DG_{AB}(\lambda_0)$, which can be described as follows:
\begin{subequations}
 \begin{align}\label{eq: contributions}
\DG_{AB}(\lambda_0) &= G_B(\lambda_0)-G_A(\lambda_0) \\
&= \DG_{AB}^0 +  \DW_{AB}^{\rm handles} + \DW_{AB}^{\rm bead} + \DW_{AB}^{\rm ssDNA} + \DW_{AB}^{\rm d}.
\end{align}
\end{subequations}
Here, $A$ ($B$) stands for the configuration of the hairpin at the beginning (ending) of the cyclic protocol at $\lambda_0$. In what follows, $f_A$ ($f_B$) is the force acting on the molecular setup when the hairpin is in state $A$ ($B$) at $\lambda_0$.  

The term $\DG_{AB}^0=G_B^0-G_A^0$ is the difference between the free energy of formation of the conformations of the DNA hairpin in states $A$ and $B$. This term depends on the sequence of the hairpin and is usually calculated using the nearest-neighbor model and the unified oligonucleotide set of basepair free energies \cite{huguet2010single,mfold} or can be recovered from pulling experiments performed in the absence of binding agents using fluctuation relations \cite{collin2005verification}. 

The two terms $\DW_{AB}^{\rm handles}$ and $\DW_{AB}^{\rm bead}$ correspond to the reversible work needed to stretch the handles and move the bead captured in the optical trap from state $A$ to state $B$. For short handles: 
\begin{equation}
 \DW_{AB}^{\rm handles} + \DW_{AB}^{\rm bead} = \frac{f^2_B-f^2_A}{2k_{\rm eff}},
\end{equation}
where $k_{\rm eff}$ is the effective stiffness of the experimental setup, equal to the slope of the force-distance curve measured in the force-branch corresponding to the native state of the hairpin. For long handles:
\begin{align}
 \label{eq: DWhOT}
 \DW_{AB}^{\rm handles} + \DW_{AB}^{\rm bead} &= \int_{x_{\rm h}(f_A)}^{x_{\rm h}(f_B)} f(x')dx' + \frac{f^2_B-f^2_A}{2k_b},
\end{align}
where $x_{\rm h}(f_A)$ ($x_{\rm h}(f_B)$) is the equilibrium end-to-end distance of the handles at force $f_A$ ($f_B$), which is calculated according to the worm-like chain model using a persistence length equal to 43.7~nm, a contour length equal to 446.08~nm and a Young modulus of 1280~pN \cite{joanthio}; 
and $k_b=0.068$~pN/nm is the stiffness of the optical trap in our setup \cite{forns2011improving}. 

The term $\DW_{AB}^{\rm ssDNA}=W_{B}^{\rm ssDNA}-W_{A}^{\rm ssDNA}$ corresponds to the difference between the reversible work needed to stretch the released single stranded DNA in configurations $B$ and $A$ from zero force to $f_A$ and $f_B$, respectively. This is calculated according to:
\begin{equation}
 \label{eq: DWss}
 \DW_{AB}^{\rm ssDNA} = \int_0^{x_{\rm ssDNA}(f_B)}f(x')dx' - \int_0^{x_{\rm ssDNA}(f_A)}f(x')dx',
\end{equation}
where equilibrium relation between the force and the end-to-end distance $f(x)$ and its inverse $x_{\rm ssDNA}(f)$ are modeled according to the worm-like chain ideal elastic model with a persistence length equal to 1.35~nm and an inter-phosphate distance of 0.59~nm/base, and the number of bases released as single-stranded DNA depends on state $B$ or $A$ \cite{alemany14}. 

The term $\DW_{AB}^{\rm d}$ is the difference between reversible work needed to orient the double-helix diameter between states $A$ and $B$:
\begin{equation}
 \label{eq: DWd}
 \DW_{AB}^{\rm d} = \int_0^{x_{\rm d}(f_B)}f(x')dx' - \int_0^{x_{\rm d}(f_A)}f(x')dx'.
\end{equation}
The helix diameter is modeled as a single bond of length $d=2$~nm that is oriented due to the action of an external force $f$ \cite{alemany14,forns2011improving}. 

\subsection{Equilibrium experiments for the hairpin-oligonucleotide system}

In equilibrium experiments in passive-mode the position of the optical trap is held constant. Hairpin ``Oligo'' (section \ref{sec: Summary of DNA hairpins used in this work}) hops rapidly between the unfolded (low forces) and the folded (high forces) states (Fig.~\ref{fig: equilibrium expts}a)~\cite{forns2011improving}. The binding and unbinding of the 10-bp oligonucleotide to its complementary sequence in the hairpin occur at a slower timescale (several seconds), and consequently binding/unbinding events can be readily identified in experiments in which a concentration of binding oligonucleotide is present (Fig.~\ref{fig: equilibrium expts}b).

To extract the binding free energy of the oligo to the unfolded DNA hairpin we consider the reaction pathway $N \leftrightarrows U \leftrightarrows B$, 
where $N$ corresponds to the state where the hairpin is in its native state, $U$ corresponds to the state where the hairpin is unfolded (and the oligo is not bound), and $B$ corresponds to the state where the hairpin is unfolded and an oligo bound. 
Since in equilibrium experiments we cannot distinguish between states $U$ and $B$ due to the very similar extension of dsDNA and ssDNA at the relevant range of forces for these experiments (Fig.~\ref{fig: equilibrium expts}b)~\cite{comstock2011ultrahigh}, we define the joint probability $\rho_{UB}=\rho_U+\rho_B$.
By considering that detailed balance is verified, it can be shown that: 
\begin{equation}
 \beta\DGb = \log \left(\frac{\rho_{UB}}{\rho_N}e^{\beta\DG_{NU}}-1\right) + \beta\DG_{UB}.
\end{equation}

In Fig.~\ref{fig: pf eq} we show 
force-time trajectories with the corresponding experimental probability density functions obtained at different values of $\lambda$ at 400~nM~oligo. From the fit to a double Gaussian (blue dashed line) we can extract the weights $\rho_N$ and $\rho_{UB}$, and determine the two forces levels ($f_N$ and $f_U=f_B$) as the average force of each Gaussian peak. 
In Table~\ref{tab: equilibrium expts} we summarize the different contributions to extract $\DGb$ for three different experimental traces. Since $f_U=f_B$, the terms $\Delta W_{UB}^{\rm handles}$, $\Delta W_{UB}^{\rm bead}$ and $\Delta W_{UB}^{\rm d}$ equal zero. 

In average, we find that at a concentration of 400 nM oligo $\langle\beta\left(\mu-\e\right)\rangle_{\rm 400 nM [oligo]} = 7\pm 1$, 
in good agreement with the theoretical predictions and with the results obtained by applying the FTLB in non-equilibrium pulling experiments.

\newpage

\section{Supplementary Figures}

\addcontentsline{toc}{subsection}{Fig. S1: Additional force-distance curves of EcoRI binding to DNA }
\begin{figure}[h!]
 \centering
\includegraphics[scale=0.8]{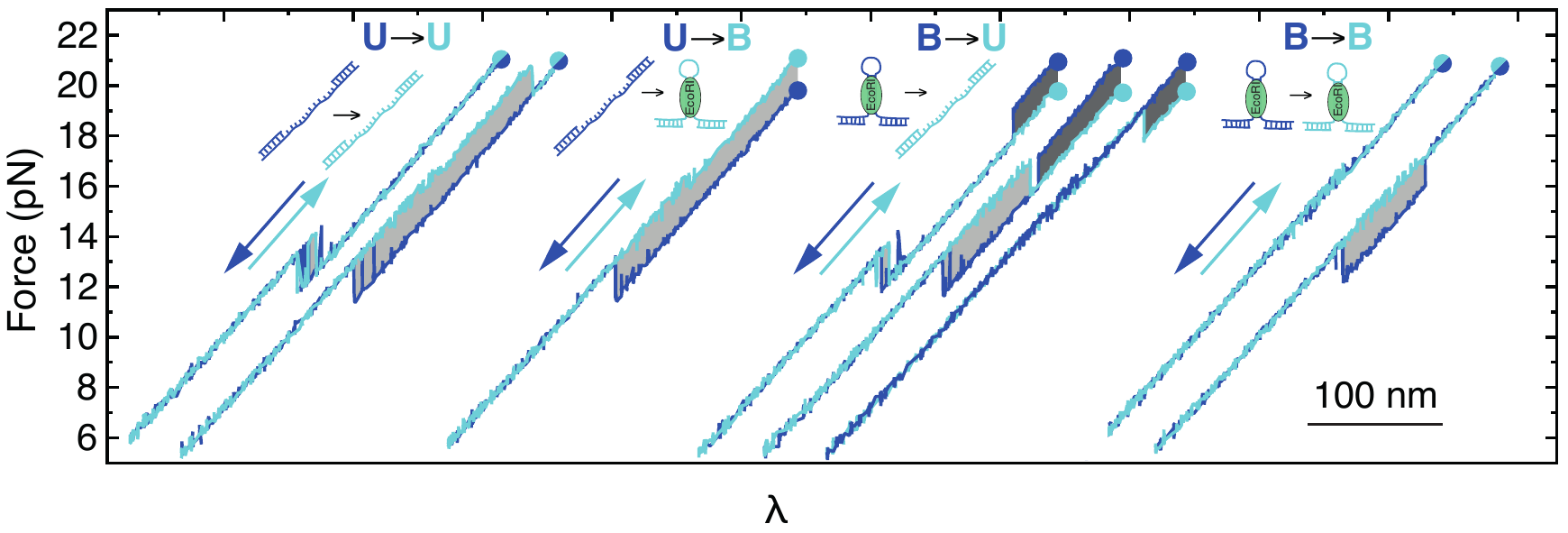}
\caption{{\bf Force-distance curves of EcoRI binding to DNA.} Example of cyclic pulling curves classified according to their initial (blue dot) and final state (cyan dot) that start and end at a high force ($\sim$21 pN). Work is calculated by integrating the area between the two curves and is shown in dark/light gray for positive/negative work values.}
\label{ecori_1}
\end{figure}

\clearpage
\addcontentsline{toc}{subsection}{Fig. S2: Effect of the upper unfolding force on the binding free energy}
\begin{figure}[h!]
 \centering
\includegraphics[scale=1]{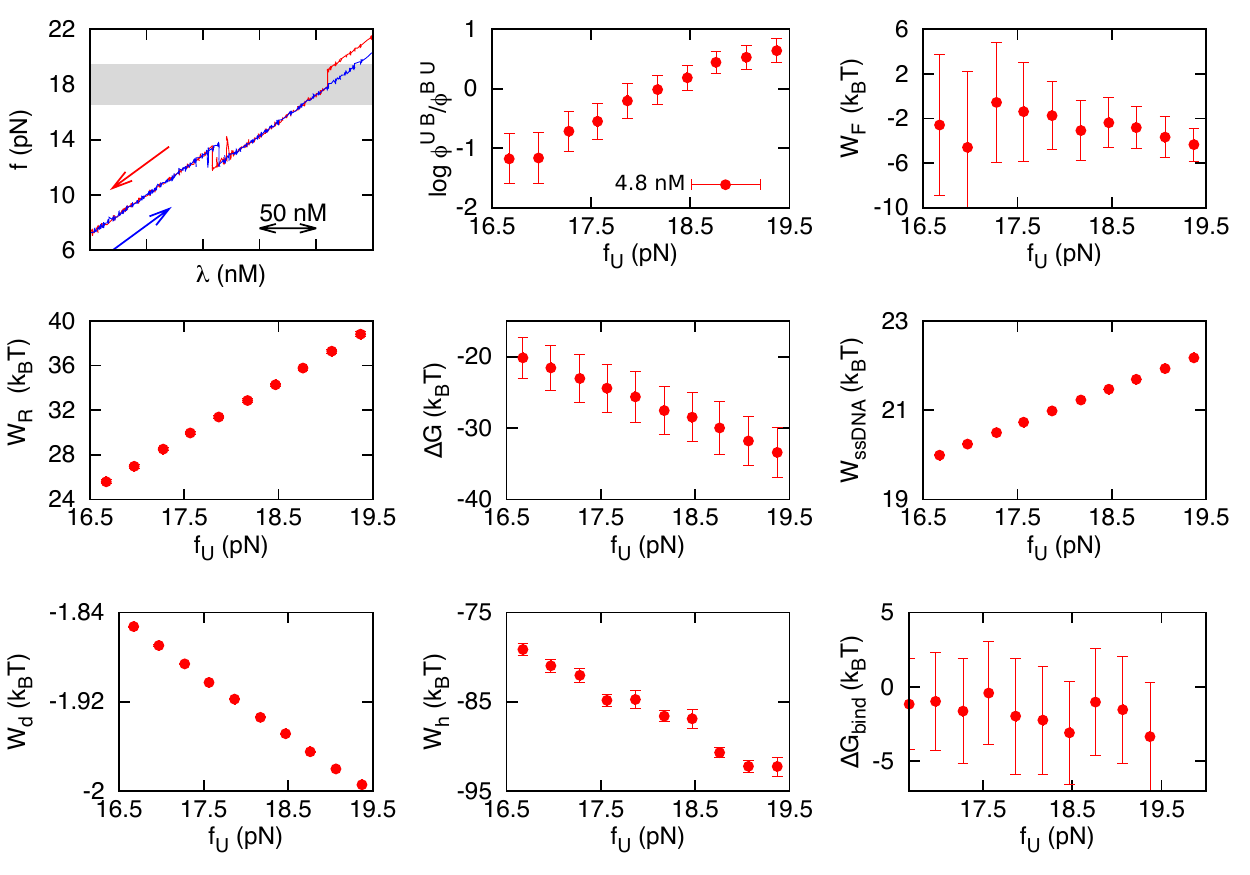}
\caption{{\bf Effect of the upper unfolding force on the derivation of the binding free energy.} The intitial/final value of $\lambda$ can be set to any position where we can unambiguously distinguish states $B$ and $U$. For each value of $\lambda$, a corresponding force $f_U$ is observed in the unfolded branch (gray area in top-left panel). We then apply the FTLB to extract the binding energy of EcoRI for different positions of $\lambda$ at high forces. The rest of the panels show the dependence on force $f_U$ of the prefactor $\log \left(\phi^{U\to B}/\phi^{B\to U}\right)$; the forward and reversed mean work values $\langle W_F\rangle$ and $\langle W_R\rangle$; the free energy difference recovered with the direct application of the FTLB, $\Delta G$; the contribution to $\Delta G$ due to the released ssDNA, the hairpin diameter and the handles of the system $W_{\rm{ssDNA}}$, $W_{\rm{d}}$ and $W_{\rm {h}}$ respectively; and finally, the free energy of binding $\DGb$. It can be seen that the resulting value of $\DGb$ does not depend on $f_U$, therefore does not depend on the initial/final position of $\lambda$ in the cyclic pulling protocol. Data shown corresponds to experiments performed at 4.8 nM for one molecule where  a total of 413 cycles where recorded. 
}
\label{fig: ecoRI force effect}
\end{figure}

\clearpage
\addcontentsline{toc}{subsection}{Fig. S3: Effect of the miscategorizing on the binding free energy}
\begin{figure}[h!]
 \centering
\includegraphics[scale=0.8]{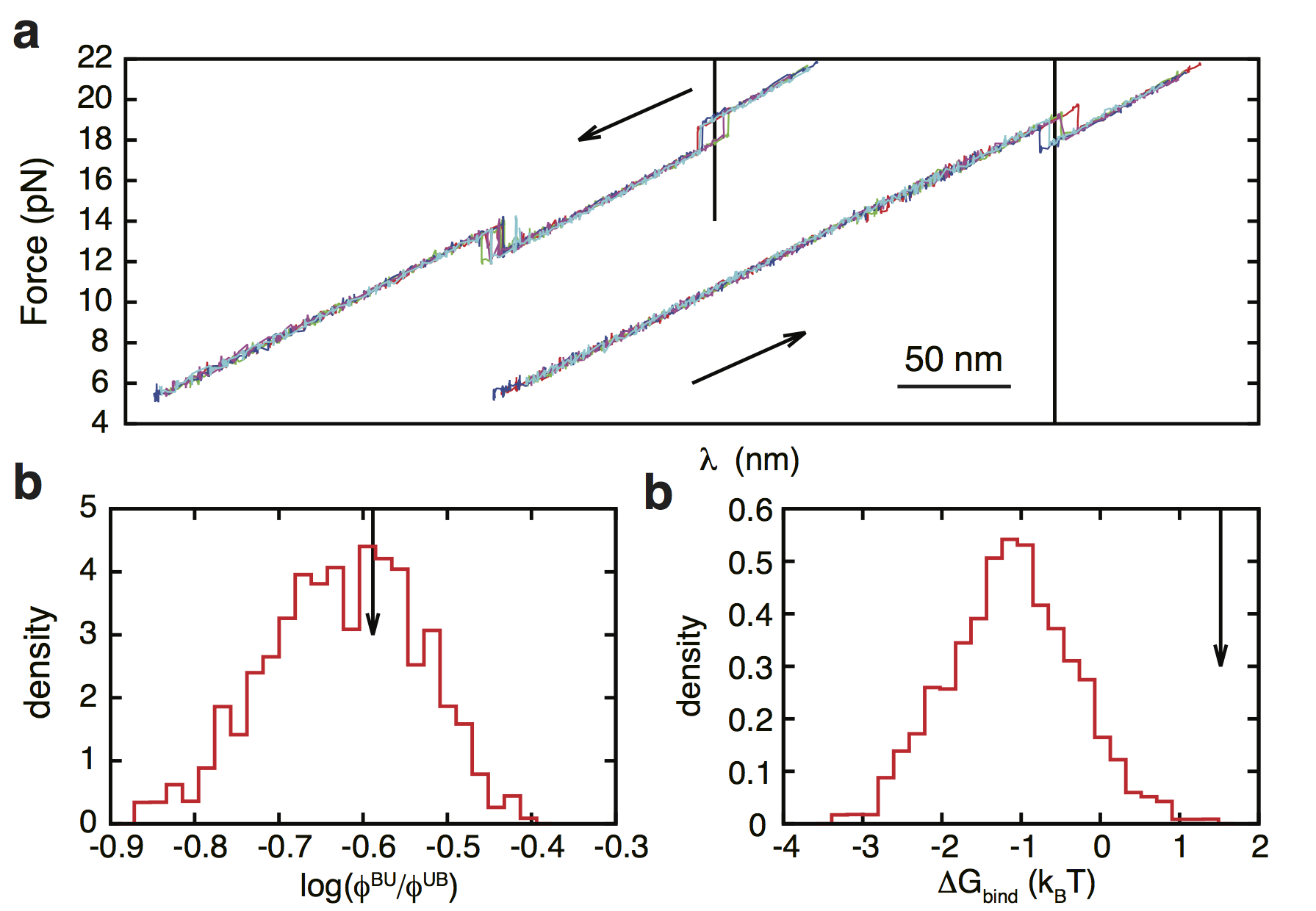}
\caption{{\bf Effect of random classification of B, U states in transitions close to the initial/final $\lambda$.} \textbf{(a)} Example of folding (left) and unfolding (right) force-distance curves where a transition $B\to U$ occurs close to the initial/final value of $\lambda$ (indicated with a vertical black line respectively).  Probabily density function of values for $\kBT\log\left(\phi^{U\to B}/\phi^{B\to U}\right)$ \textbf{(b)} and $\DGb$ \textbf{(c)} obtained by of randomly assigning 500 independent times states B and U to trajectories where an unbinding transition is observed at $\lambda\pm10$ nm (such as the ones depicted in panel a). Vertical arrows indicate the value recovered with the correct classification of the initial/final states along the cyclic pulling protocol. We observed that a random classification of states in trajectories were transitions are observed close the the inifial/final value of $\lambda$ persistently leads to lower values for the binding free energy. 
}\label{fig: ecoRI random}
\end{figure}

\clearpage
\addcontentsline{toc}{subsection}{Fig. S4: Equilibrium experiments of oligonucleotide binding}
\begin{figure}[h]
 \centering
\includegraphics[scale=.7]{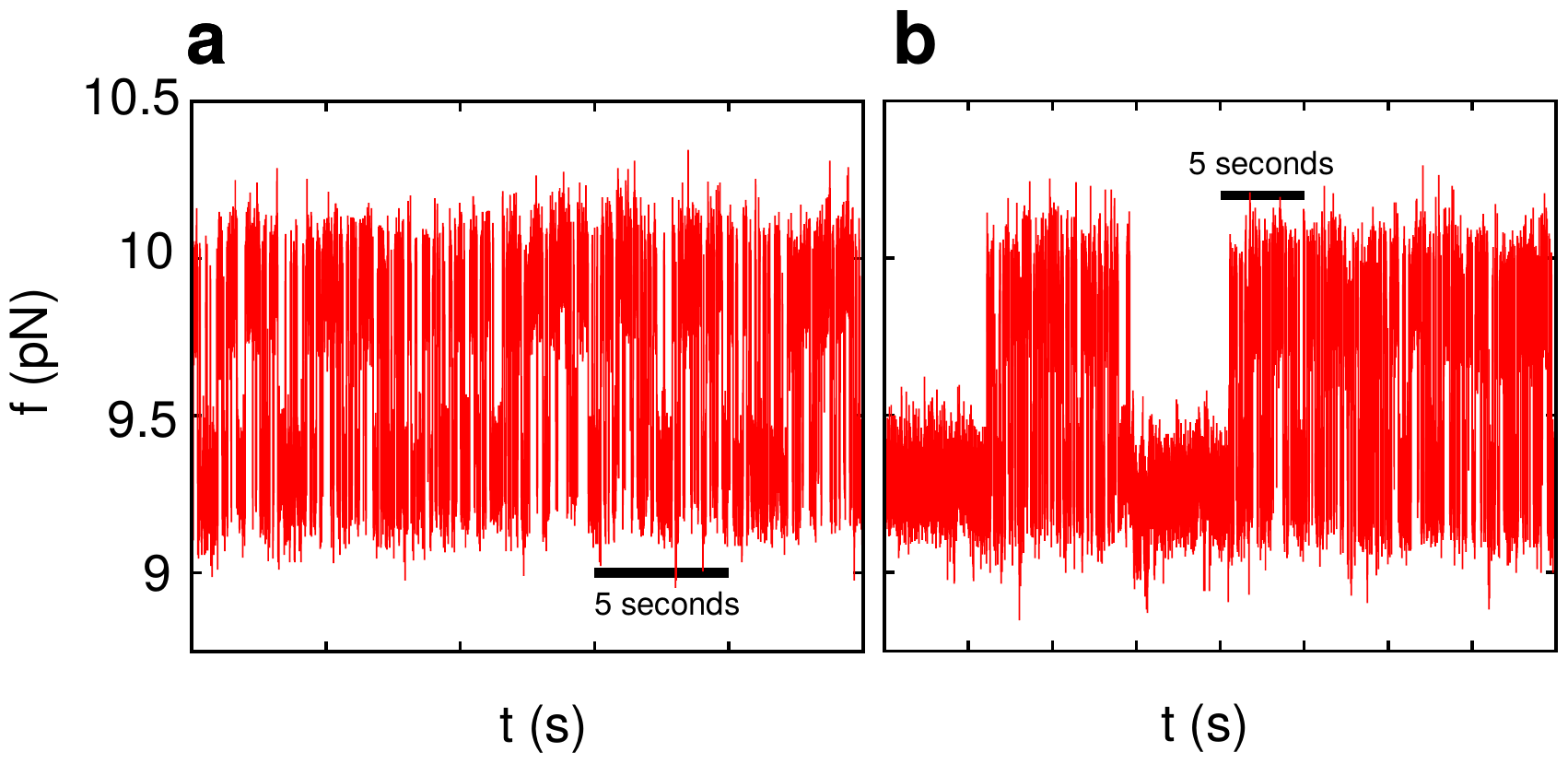}
\caption{{\bf Experimental traces of equilibrium experiments of oligonucleotide binding.} 
{\bf (a)}~Equilibrium experiments performed without oligo with hairpin ``Oligo''. 
{\bf (b)}~Equilibrium experiments performed at 400~nM~[oligo]. Two time-scales are revealed when the hairpin is in the unfolded state (low forces).}\label{fig: equilibrium expts}
\end{figure}

\clearpage

\addcontentsline{toc}{subsection}{Fig. S5: Equilibrium experiments at different forces}
\begin{figure}[h]
 \centering
\includegraphics[scale=0.55]{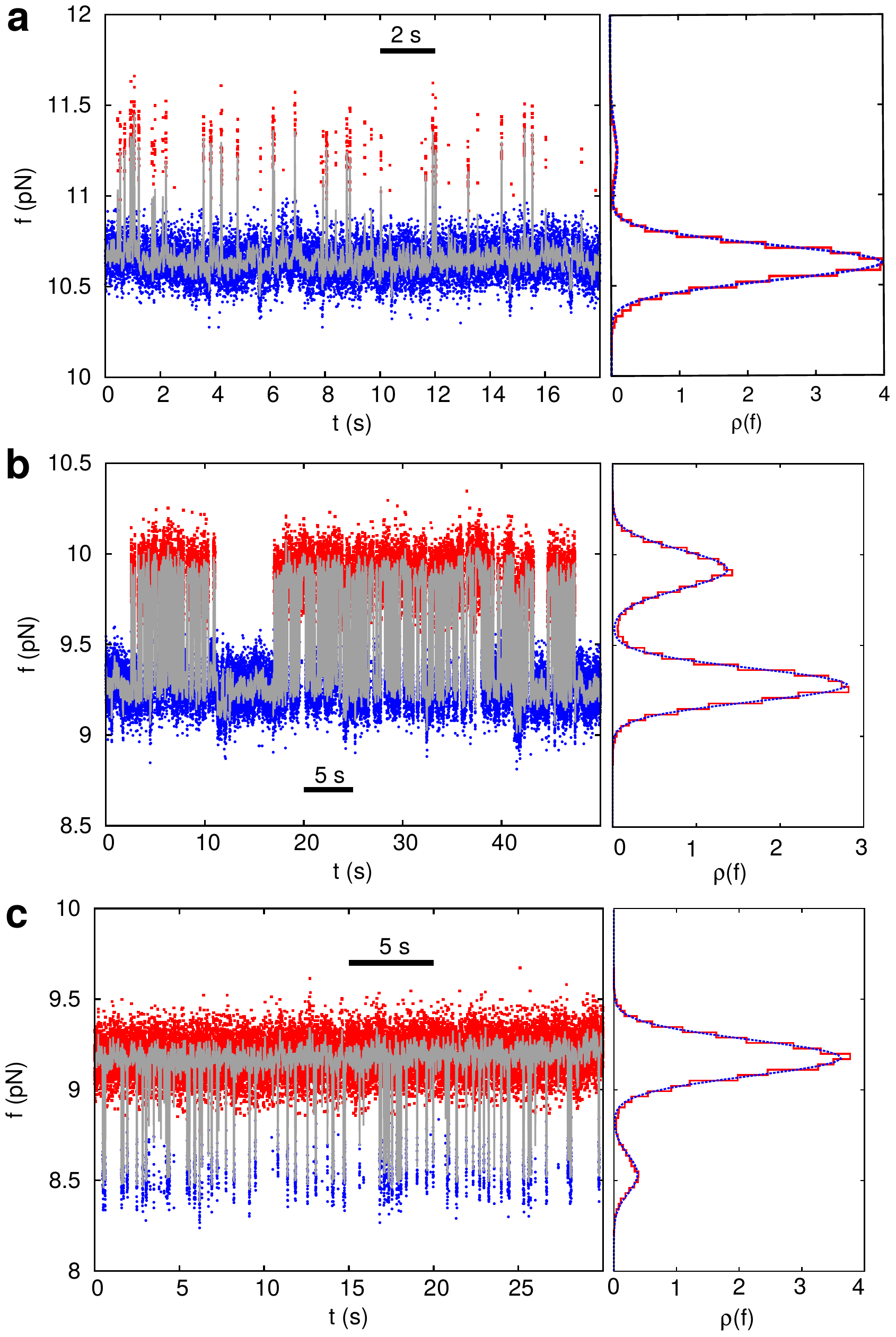}
\caption{{\bf Equilibrium experiments at different forces.} In gray we show an averaged experimental trace; in red we highlight experimental data points (acquisition rate: 1 kHz) where the hairpin is in the folded state (and therefore no oligo is bound), whereas in blue we highlight data points where the hairpin is in the unfolded state (either with the oligo bound or not bound). Two time-scales are observed in the experiments showing that the oligo binds and unbinds from the hairpin in an stochastic manner and with a timescale much longer than the folding/unfolding rate of the hairpin. The panels on the right show an histogram of the probability density (red) and a double gaussian fit to the data (blue).}\label{fig: pf eq}
\end{figure}

\clearpage
\addcontentsline{toc}{subsection}{Fig. S6: Theoretical free energy branches for oligo binding}
\begin{figure}[h]
 \centering
\includegraphics[scale=1]{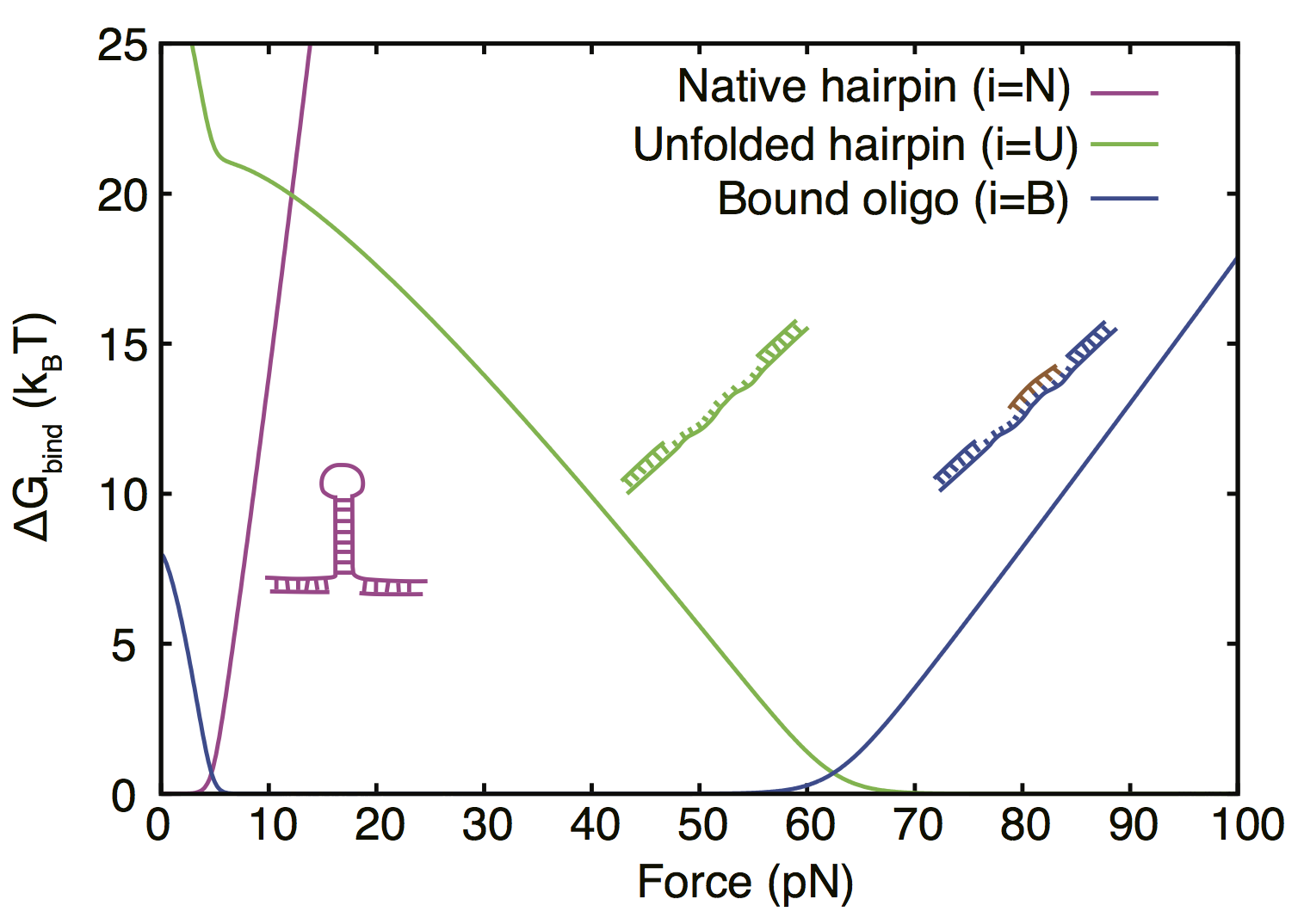}
\caption{{\bf  Free energy of branches $B$, $N$ and $U$.} Free energy branches are shown as a function of force and computed relative to the total free energy of the system (also called potential of mean force, equal to $-\log\left[\exp(-\beta\Delta G_B) + \exp(-\beta\Delta G_N) + \exp(-\beta\Delta G_U)\right]$). The free energy of state $B$ is computed by taking into account the free energy of binding (equal to 20 $\kBT$) plus the elastic response of a DNA made of two ligated ssDNA and dsDNA segments. The first segment is a 24 bases-long ssDNA chain (modeled with the worm-like chain (WLC) model with persistence length equal to 1.5 nm and inter-phosphate distance equal to 0.59 nm/base). The second segment is the elastic response of a 10 basepair-long dsDNA chain (modeled with the WLC model with persistence length equal to 50 nm and inter-phosphate distance equal to 0.34 nm/base). The free energy of state $U$ is computed by taking only into account the elastic response of a 34 bases-long ssDNA chain. Finally, the free energy of state N contains the folding free energy of the hairpin (28 $\kBT$) plus the elastic response of the hairpin diameter, modeled as a bond of length 2.0 nm (equal to the hairpin diameter) that is oriented in the presence of a force. It can be seen that at low forces state $N$ is the most stable. However, at $\sim$5 pN state $B$ becomes the most stable until $\sim$62 pN, where state $U$ becomes more stable. Therefore, with this simple model we predict that the threshold force above which the oligo will not bind is $\sim$62 pN. However, at those high forces potential perturbations of the force into the ssDNA structure neglected in the model might change this value. }
\end{figure}

\clearpage

\addcontentsline{toc}{subsection}{Fig. S7: Force-distance curves of hairpins SP and NSP}
\begin{figure}
 \centering
\includegraphics[scale=0.8]{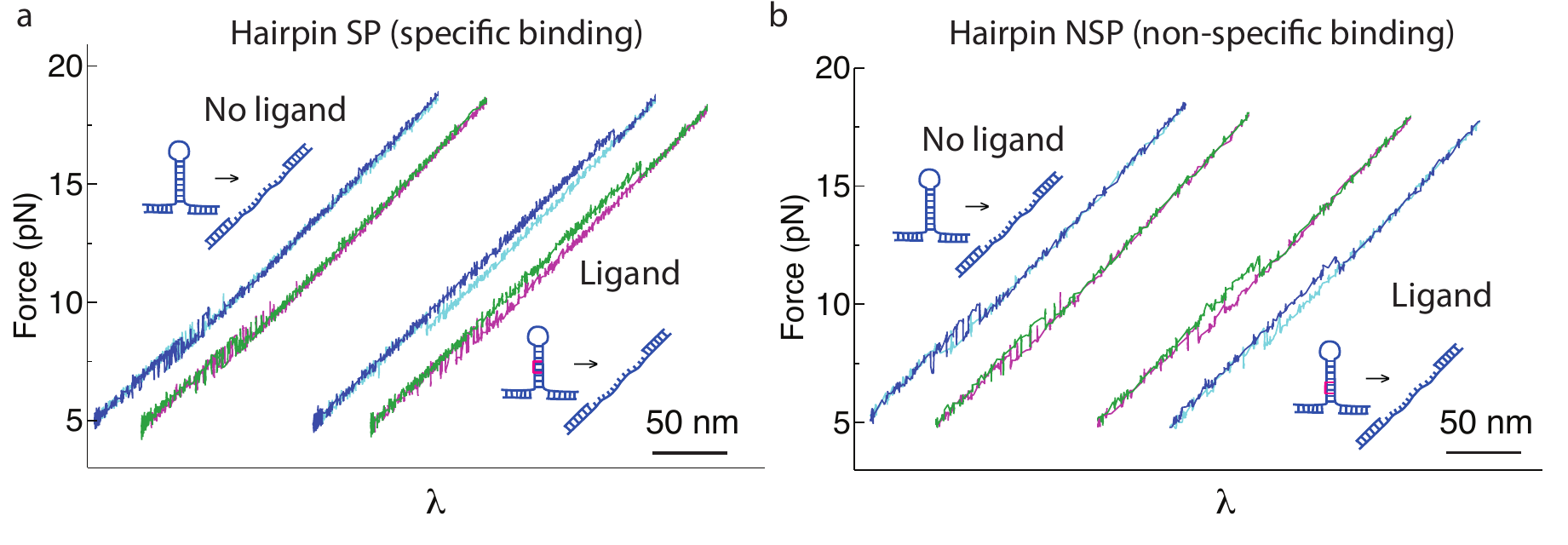}
\caption{{\bf Force-distance curves of hairpins SP and NSP.} {\bf(a)} FDCs of hairpin SP in the absence (left) and presence of ligand (right). {\bf(b)} FDCs of hairpin NSP in the absence (left) and presence of ligand (right). In each FDC blue/green is unfolding and cyan/magenta refolding. Pulling speed is 70 nm/s in (a) and 250 nm/s in (b).}
\label{echi}
\end{figure}

\clearpage

\addcontentsline{toc}{subsection}{Fig. S8: Direct measurement of binding kinetic rates for echinomycin in hairpin SP}
\begin{figure}
 \centering
\includegraphics[scale=0.8]{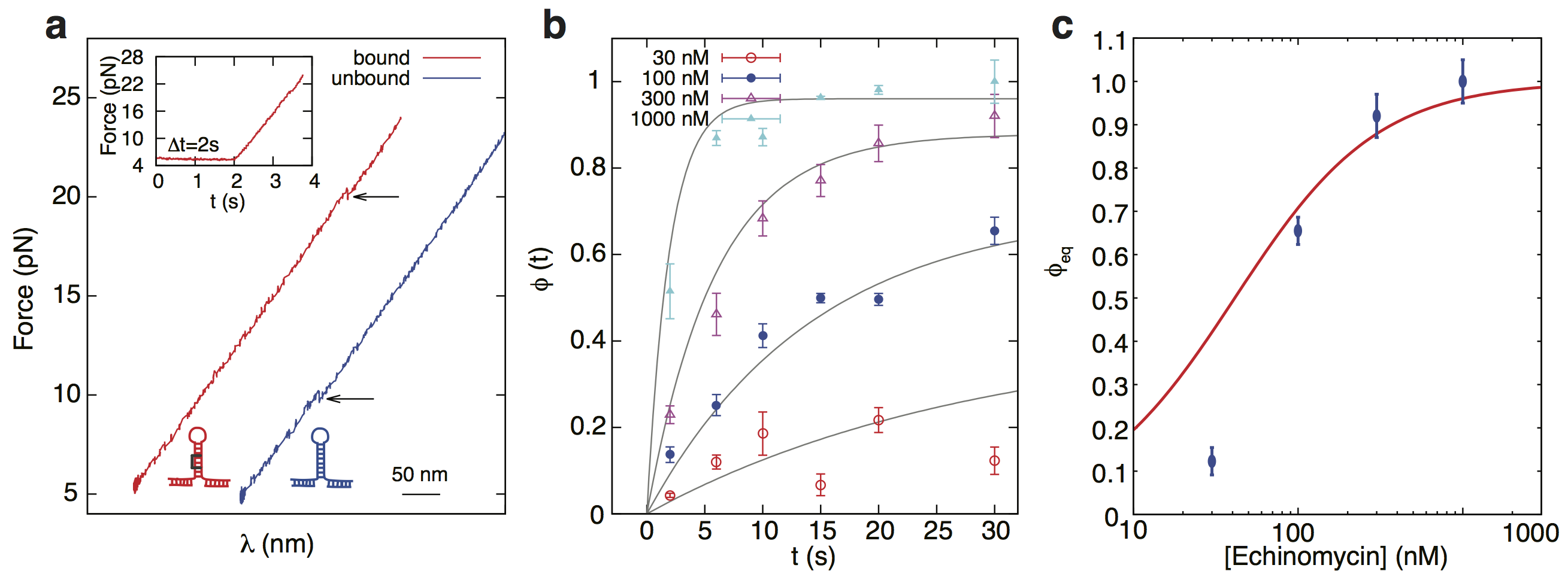}
\caption{{\bf Specific ligand binding of echinomycin to DNA.} {\bf(a)} Fast mechanical unfolding is performed after incubating the hairpin at $\sim$5 pN during a fixed time interval $\Delta t$. The unfolding force of the hairpin, indicated with an arrow, relates to the bound state of echinomycin: unfolding forces above (below) 15 pN (red(blue) curve) indicate that an echinomycin molecules has (has not) bound to the hairpin (Fig. 3a, main document). 
{\bf (b)} Fraction of bound states as a function of the time $\Delta t$ and concentration of echinomycin and fit to  to a first order reaction kinetics model (DNA + I$\rightleftarrows$DNA$\cdot$I) where $\phi(t) = \frac{k_\to [I]}{k_\leftarrow [I] + k_\leftarrow}\left(1-\exp[(k_\to[I] + k_\leftarrow)t]\right)$. From the fit, we obtain 
$k_\to=(4.9\pm0.4)\times 10^{-4}$ nM$^{-1}$s$^{-1}$
and $k_\leftarrow=(2.0\pm 0.5)\times 10^{-2}$ s$^{-1}$, which implies $K_d=k_\leftarrow/k_\to= 41 \pm 10$ nM and $\DGb=17\pm1$ $\kBT$. This result is in good agreement with the value obtained using the FTLB ($\DGb=20\pm1$ $\kBT$). 
{\bf (c)} Binding isotherm of echinomycin determined from optical trapping. The red curve has been obtained from the fit in panel b. Blue points are the fraction of bound population measured as described in panel a at the largest measured time $\Delta t=30$s. The disagreement between theory and experiments observed at low concentrations shows that binding kinetics is still out of equilibrium at the largest measured time of 30s. Error bars are standard errors computed by averaging over different molecules.
}
\label{fig: echi titration}
\end{figure}

\clearpage

\addcontentsline{toc}{subsection}{Fig. S9: Force-distance curves of hairpins C and NC}
\begin{figure}
 \centering
\includegraphics[scale=0.8]{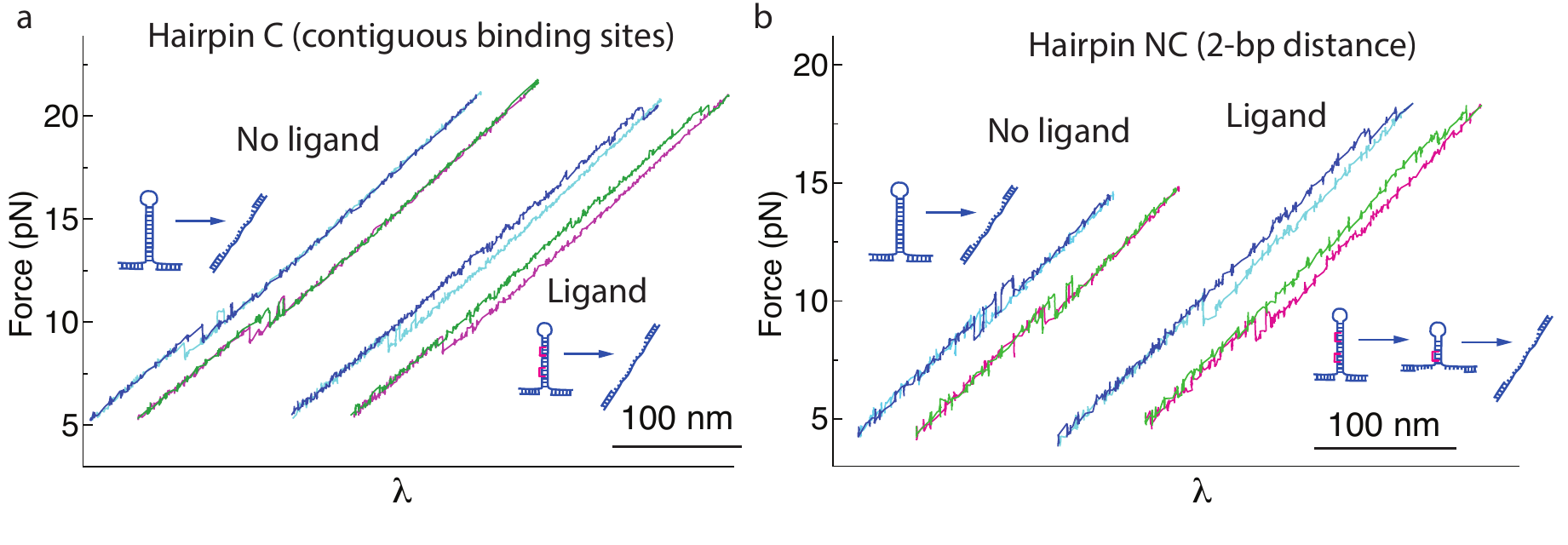}
\caption{{\bf Force-distance curves of hairpins C and NC.} {\bf(a)} FDCs of hairpin C in the absence (left) and presence of ligand (right) {\bf(b)} FDCs of hairpin NC in the absence (left) and presence of ligand.  In the presence of ligand, FDCs of hairpin C show higher unfolding forces than those of hairpin NC, in agreement with the proposed cooperative effect between ligand pairs. Similarly a partially unfolded intermediate with just one ligand bound is observed in hairpin NC, whereas hairpin C cooperatively unfolds in a single step. In each FDC blue/green is unfolding and cyan/magenta refolding. Pulling speed is 70 nm/s in (a) and 250 nm/s in (b).}   
\label{coop}
\end{figure}

\clearpage

\addcontentsline{toc}{subsection}{Fig. S10: Force-distance curves of hairpin M at [Echinomycin]=10 $\mu$M}
\begin{figure}
 \centering
\includegraphics[scale=1.1]{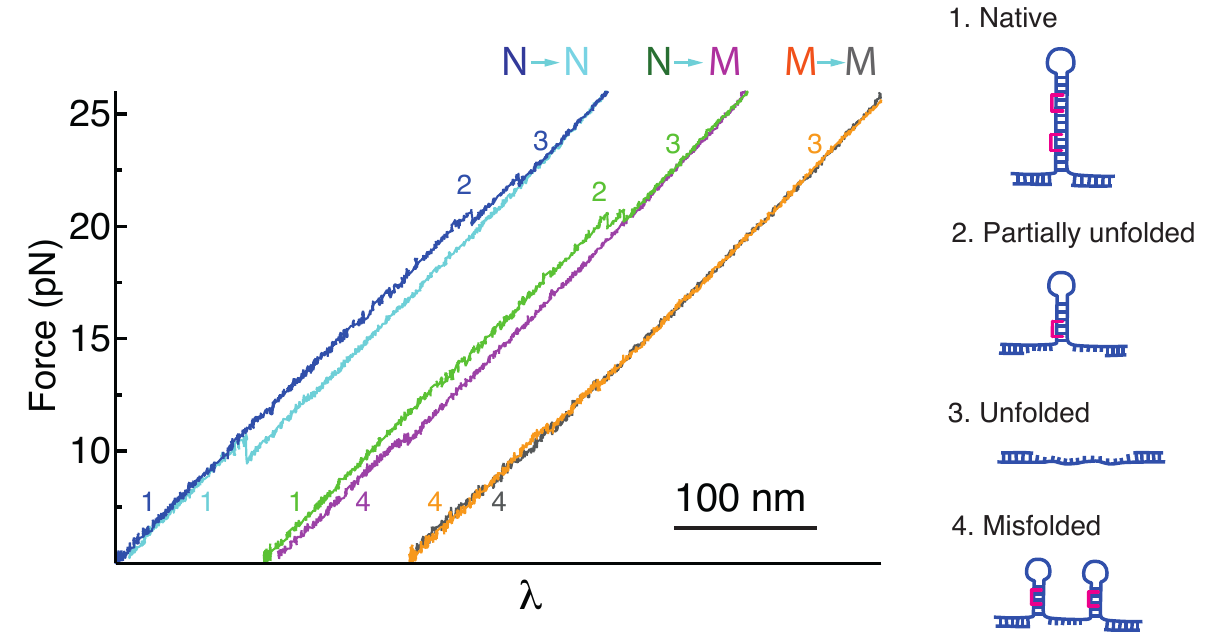}
\caption{{\bf Force-distance curves of hairpin M at [Echinomycin]=10 $\mu$M.}  The ligand can kinetically trap a misfolded state consisting of two 4-bp DNA hairpins serially connected. Characteristic pulling curves connecting the native (N) and misfolded (M) state are shown. The different molecular configurations observed during the pulling curve are indicated in the scheme. Pulling speed is 70 nm/s.}
\label{misfolding}
\end{figure}

\clearpage

\section{Supplementary Tables}

\addcontentsline{toc}{subsection}{Tab. S1: Number of experiments performed at 130~mM NaCl for different concentrations of EcoRI}
\begin{table}[h!]
\begin{tabular}{lll}
\hline\hline
[EcoRI] (nM) & pulled molecules (number of cycles per molecule) & total cycles $N$ \\
 \hline
0.25 & 5 (141, 87, 95, 360, 101) & 784\\
0.50 & 8 (297, 93, 43, 101, 40, 18, 137, 27) & 756\\
1.00 & 8 (852, 555, 486, 48, 153, 76, 159, 404) & 2733\\
2.40 & 8 (170, 356, 350, 433, 212, 93, 245, 126) & 1985\\
4.80 & 3 (413, 339, 349) & 1101 \\
10.0 & 10 (470, 317, 290, 242, 87, 200, 248, 401, 102, 396) & 2753 \\
20.0 & 3 (257, 242, 668) & 1167\\
\hline
\end{tabular}
\caption{{\bf Number of experiments performed at 130~mM NaCl for different concentrations of EcoRI.} Number of molecules measured at each concentration of EcoRI, corresponding cycles per molecule shown in parenthesis, and total number of cycles used for computing the binding energy.}\label{tab: EcoRI N}
\end{table}

\newpage

\addcontentsline{toc}{subsection}{Tab. S2: Contributions to the binding free energy of EcoRI to dsDNA as a function of [EcoRI] at 130 mM NaCl}
\begin{table}
\begin{tabular}{cccccc}
\hline\hline
$f_B$ (pN) & $f_U$ (pN) & $\DG_{nU}^0$  & $\DW_{nU}^{\rm handles}+\DW_{nU}^{\rm bead}$ & $\DW_{nU}^{\rm ssDNA}$ & $\DW_{nU}^d$ \\
 \hline
 19.77$\pm$0.01 & 18.54$\pm$0.01 & 42$\pm$2 & -92$\pm$1 & 21.51$\pm$0.01 & -1.955$\pm$0.001\\
\end{tabular}
\begin{tabular}{cccccc}
\hline\hline
 [EcoRI] (nM) & $\phi^{B\to U}$ & $\phi^{U\to B}$ & $\log \left(\phi^{B\to U}/\phi^{U\to B}\right)$ & $\DG_{BU}$ & $\DGb$ \\
 \hline
 0.25 & 0.4$\pm$0.1   & 0.12$\pm$0.03 & 1.2$\pm$0.3  & -29$\pm$2 & 2$\pm$4 \\
 0.50 & 0.24$\pm$0.05 & 0.22$\pm$0.02 & 0.1$\pm$0.2  & -25$\pm$2 & 5$\pm$4 \\
 1.00 & 0.15$\pm$0.03 & 0.31$\pm$0.02 & -0.8$\pm$0.1 & -23$\pm$1 & 5$\pm$3 \\
 2.40 & 0.24$\pm$0.04 & 0.26$\pm$0.03 & -0.2$\pm$0.2 & -24$\pm$2 & 8$\pm$2 \\
 4.80 & 0.24$\pm$0.04 & 0.27$\pm$0.03 & -0.1$\pm$0.2 & -27$\pm$2 & 5$\pm$3 \\
 10.0 & 0.17$\pm$0.02 & 0.71$\pm$0.05 & -1.5$\pm$0.1 & -22$\pm$1 & 8$\pm$1 \\
 20.0 & 0.13$\pm$0.02 & 0.83$\pm$0.03 & -1.9$\pm$0.1 & -23$\pm$1 & 8$\pm$1 \\
 \hline
\end{tabular}
\caption{{\bf Contributions to the binding free energy of EcoRI to dsDNA as a function of [EcoRI] at 130 mM NaCl.}
For all the pulling experiments performed for different molecules at different concentrations of EcoRI, the initial/final value of the control parameter $\lambda$ was chosen so that forces $f_B$ and $f_U$ are on average the same. Hence, numerical values for $\DG_{nU}^0$, $\DW_{nU}^{\rm handles}+\DW_{nU}^{\rm bead}$ (Eq.~\ref{eq: DWhOT}),  $\DW_{nU}^{\rm ssDNA}$ (Eq.~\ref{eq: DWss}), and $\DW_{nU}^{\rm d}$ (Eq.~\ref{eq: DWd}) are also on average the same for different molecules pulled at different concentrations of EcoRI. In contrast, $\phi^{B\to U}$, $\phi^{U\to B}$, $\log \left(\phi^{B\to U}/\phi^{U\to B}\right)$, $\DG_{BU}$ and $\DGb$ depend on the concentration of ligand. Here $n=7$.
}\label{tab: EcoRI 1}
\end{table}

\clearpage

\addcontentsline{toc}{subsection}{Tab. S3: Number of experiments performed at 1~nM EcoRI for different concentrations of NaCl}
\begin{table}
\begin{tabular}{lll}
\hline\hline
[NaCl] (mM) & pulled molecules (number of cycles per molecule) & total cycles $N$ \\
 \hline
 60  & 15 (156, 69, 525, 60, 61, 35, 21, 43, 713, 258, 433, 58, 78, 162, 83) & 2755\\
 75  & 6 (263, 312, 519, 506, 156, 302) & 2058 \\
 100 & 7 (276, 273, 516, 577, 128, 494, 235) & 2499 \\
 130 & 8 (852, 555, 486, 48, 153, 76, 159, 404) & 2733 \\
 180 & 8 (654, 323, 88, 298, 61, 423, 211, 80) & 2138 \\
 \hline
\end{tabular}
\caption{{\bf Number of experiments performed at 1~nM EcoRI for different concentrations of NaCl.}}\label{tab: ecori N sal}
\end{table}

\clearpage

\addcontentsline{toc}{subsection}{Tab. S4: Contributions to the binding free energy of EcoRI to dsDNA as a function of [NaCl] at 1~nM EcoRI}

\begin{table}
\begin{tabular}{ccccc}
\hline\hline
 $f_B$ (pN) & $f_U$ (pN) & $\DW_{nU}^{\rm handles}+\DW_{nU}^{\rm bead}$ & $\DW_{nU}^{\rm ssDNA}$ & $\DW_{nU}^d$ \\
 \hline
 19.76$\pm$0.01 & 18.56$\pm$0.01 &  -91$\pm$1 & 21.54$\pm$0.01 & -1.955$\pm$0.001\\
\end{tabular}
\begin{tabular}{ccccccc}
\hline\hline
 [EcoRI] (nM) & $\DG_{nU}^0$ & $\phi^{B\to U}$ & $\phi^{U\to B}$ & $\log \left(\phi^{B\to U}/\phi^{U\to B}\right)$ & $\DG_{BU}$ & $\DGb$ \\
 \hline
 60  & 39$\pm$2 & 0.056$\pm$0.008 & 0.8$\pm$0.1   & -2.5$\pm$0.4 & -19$\pm$3 & 13$\pm$3 \\
 75  & 40$\pm$2 & 0.06$\pm$0.01   & 0.7$\pm$0.1   & -2.5$\pm$0.4 & -20$\pm$1 & 11$\pm$2 \\
 100 & 41$\pm$2 & 0.14$\pm$0.02   & 0.29$\pm$0.05 & -0.9$\pm$0.2 & -25$\pm$2 &  5$\pm$2 \\
 130 & 42$\pm$2 & 0.15$\pm$0.03   & 0.31$\pm$0.02 & -0.8$\pm$0.1 & -23$\pm$1 &  6$\pm$1 \\
 180 & 44$\pm$2 & 0.44$\pm$0.07   & 0.06$\pm$0.02 &  1.6$\pm$0.4 & -27$\pm$2 &  0$\pm$2 \\
 \hline
\end{tabular}
\caption{{\bf Contributions to the binding free energy of EcoRI to dsDNA as a function of [NaCl] at 1~nM EcoRI.} 
Caption as in Table~\ref{tab: EcoRI 1}.}\label{eq: ecori g nacl}
\end{table}

\clearpage

\addcontentsline{toc}{subsection}{Tab. S5: Number of experiments performed at different concentrations of oligo}
\begin{table}
\begin{tabular}{lll}
\hline\hline
[oligo] (nM) & pulled molecules (number of cycles per molecule) & total cycles $N$\\
 \hline
25   & 2 (25, 244) & 269 \\ 
50   & 7 (57, 131, 72, 242, 311, 98 46) & 957 \\ 
100  & 6 (182, 198, 22, 435, 368, 386) & 1591\\ 
200  & 5 (28, 17, 59, 67, 87) & 258\\ 
400  & 4 (264, 123, 213, 109) & 709 \\ 
1000 & 10 (59, 56, 288, 293, 325, 224, 83, 432, 317) & 2077\\
2000 & 5 (185, 603, 74, 273, 147) & 1282\\
 \hline
\end{tabular}
\caption{{\bf Number of experiments performed at different concentrations of oligo.}}\label{tab: oligo N}
\end{table}

\clearpage

\addcontentsline{toc}{subsection}{Tab. S6: Contributions to the binding free energy of oligo to complementary ssDNA as a function of the concentration of the oligo}
\begin{table}
\begin{tabular}{cccccc}
\hline\hline
 $f_B$ (pN) & $f_N$ (pN) & $\DG_{UN}^0$ & $\DW_{UN}^{\rm handles}+\DW_{UN}^{\rm bead}$ & $\DW_{UN}^{\rm ssDNA}$ & $\DW_{UN}^d$ \\
 \hline
 6.20$\pm$0.06 & 6.80$\pm$0.06 & -27.5$\pm$0.8 & 16.5$\pm$0.5 & -4.81$\pm$0.05 & 0.896$\pm$0.008 \\
\end{tabular}
\begin{tabular}{cccccc}
\hline\hline
 [oligo] (nM) & $\phi^{B\to N}$ & $\phi^{N\to B}$ & $\log \left(\phi^{B\to N}/\phi^{N\to B}\right)$ & $\DG_{BN}$ & $\DGb$ \\
 \hline
 25   & 1.0$^*$       & 0.030$\pm$0.004 & 3.6$\pm$0.1   & -11.3$\pm$0.7 & 4$\pm$1 \\
 50   & 0.96$\pm$0.04 & 0.037$\pm$0.006 & 3.4$\pm$0.2   & -11.8$\pm$0.6 & 3$\pm$1 \\
 100  & 0.94$\pm$0.03 & 0.062$\pm$0.006 & 2.79$\pm$0.08 & -11.0$\pm$0.6 & 4$\pm$1 \\
 200  & 0.87$\pm$0.08 & 0.09$\pm$0.03   & 2.3$\pm$0.3   &  -9.2$\pm$0.4 & 6$\pm$1 \\
 400  & 0.86$\pm$0.06 & 0.18$\pm$0.03   & 1.6$\pm$0.3   &  -8.4$\pm$0.3 & 7$\pm$1 \\
 1000 & 0.49$\pm$0.05 & 0.45$\pm$0.04   & 0.37$\pm$0.07 &  -7.7$\pm$0.5 & 7$\pm$1 \\
 2000 & 0.37$\pm$0.02 & 0.60$\pm$0.04   & -0.5$\pm$0.1  &  -6.5$\pm$0.8 & 8$\pm$1 \\
 \hline
\end{tabular}
\caption{{\bf Contributions to the binding free energy of oligo to complementary ssDNA as a function of the concentration of the oligo.} For all the pulling experiments performed for different molecules at different concentrations of oligo, the initial/final value of the control parameter $\lambda$ was chosen so that forces $f_B$ and $f_N$ are on average the same same. Numerical values for $\DG_{UN}^0$, $\DW_{UN}^{\rm handles}+\DW_{UN}^{\rm bead}$ (Eq.~\ref{eq: DWhOT}),  $\DW_{UN}^{\rm ssDNA}$ (Eq.~\ref{eq: DWss}), and $\DW_{UN}^{\rm d}$ (Eq.~\ref{eq: DWd}) are also on average the same for different molecules pulled at different concentrations of oligo. In contrast, $\phi^{B\to N}$, $\phi^{N\to B}$, $\log \left(\phi^{B\to N}/\phi^{N\to B}\right)$, $\DG_{BN}$ and $\DGb$ de depend on the concentration of ligand.}\label{tab: ecori g}
\end{table}

\clearpage

\addcontentsline{toc}{subsection}{Tab. S7: Determination of $\beta(\mu-\e)$ in equilibrium experiments}
\begin{table}
 \centering
\begin{tabular}{cl@{$\pm$}rl@{$\pm$}rl@{$\pm$}r}
\hline\hline
trace & \multicolumn{2}{c}{$\langle f_N\rangle$ (pN)} & \multicolumn{2}{c}{$\langle f_U\rangle$ (pN)} & \multicolumn{2}{c}{$\rho_N$} \\ 
 \hline
a & 11.22 & 0.05 & 10.62 & 0.01 & 0.038 & 0.006\\
b & 9.92  & 0.01 & 9.28  & 0.01 & 0.364 & 0.004\\
c & 9.18  & 0.01 & 8.53  & 0.01 & 0.901 & 0.005\\
\hline
\end{tabular}
\begin{tabular}{r@{$\pm$}lr@{$\pm$}lr@{$\pm$}lr@{$\pm$}lr@{$\pm$}l}
\multicolumn{2}{c}{$\beta\Delta W_{NU}^{\rm handles}+\beta\Delta W_{NU}^{\rm bead}$} & \multicolumn{2}{c}{$\beta\Delta W_{NU}^{\rm ssDNA}$} & \multicolumn{2}{c}{$\beta\Delta W_{NU}^{\rm d}$} & \multicolumn{2}{c}{$\Delta W_{UB}^{\rm ssDNA}$} & \multicolumn{2}{c}{$\beta(\mu-\e)$}\\
\hline
-29 & 2 & 10.4 & 0.5 & 1.40 & 0.02 & -2.4 & 0.5 & 7.5 & 1 \\
-28 & 2 &  9.3 & 0.5 & 1.27 & 0.02 & -2.1 & 0.5 & 6 & 1 \\
-25 & 3 &  8.7 & 0.5 & 1.19 & 0.02 & -1.9 & 0.5 & 6 & 1 \\
\hline
\end{tabular}
\caption{{\bf Determination of $\beta(\mu-\e)$ in equilibrium experiments.} Results obtained at 400~nM~[oligo].}\label{tab: equilibrium expts}
\end{table}

\clearpage

\addcontentsline{toc}{subsection}{Tab. S8: Number of experiments performed at different concentrations of echinomycin with hairpin SP}
\begin{table}
\begin{tabular}{lll}
\hline\hline
 [Echi] (nM) & pulled molecules (number of cycles per molecule) & total cycles $N$ \\
 \hline
 100  & 7 (121, 408, 369, 277, 98, 484, 261) & 2018\\
 300  & 6 (496, 271, 201, 341, 600, 511) & 2420 \\
 1000 & 9 (1302, 1272, 486, 978, 420, 121, 327, 470, 196) & 5572\\
 3000 & 6 (670, 175, 1010, 399, 1268, 475) & 3997  \\
 \hline
\end{tabular}
\caption{{\bf Number of experiments performed at different concentrations of echinomycin with hairpin SP.}}\label{tab: echi N}
\end{table}

\clearpage

\addcontentsline{toc}{subsection}{Tab. S9: Contributions to the binding free energy of echinomycin to dsDNA as a function of ligand concentration [Echi]}
\begin{table}
\begin{tabular}{cccccc}
\hline\hline
 $f_B$ (pN) & $f_U$ (pN) & $\DG_{nU}^0$  & $\DW_{nU}^{\rm handles}+\DW_{nU}^{\rm bead}$ & $\DW_{nU}^{\rm ssDNA}$ & $\DW_{nU}^d$ \\
 \hline
 16.77$\pm$0.02 & 16.25$\pm$0.02 & 10.5$\pm$0.3 & -33.0$\pm$0.8 & 8.25$\pm$0.03 & -1.790$\pm$0.006 \\
\end{tabular}
\begin{tabular}{cccccc}
\hline\hline
[Echi] (nM) & $\phi^{B\to U}$ & $\phi^{U\to B}$ & $\log \left(\phi^{B\to U}/\phi^{U\to B}\right)$ & $\DG_{BU}$ & $\DGb$ \\
 \hline
100  & 0.5$\pm$0.1     & 0.046$\pm$0.005 & -2.2$\pm$0.4 & -10.8$\pm$0.4 & 3$\pm$2 \\
300  & 0.46$\pm$0.07   & 0.112$\pm$0.008 & -1.3$\pm$0.2  & -10.6$\pm$0.6 & 5$\pm$2 \\
1000 & 0.27$\pm$0.04   & 0.18$\pm$0.05   & 0.5$\pm$0.4  & -10.3$\pm$0.8 & 5$\pm$2 \\
3000 & 0.158$\pm$0.008 & 0.59$\pm$0.04   & 1.3$\pm$0.1 & -8.6 $\pm$0.6 & 10$\pm$1 \\
 \hline
\end{tabular}
\caption{{\bf Contributions to the binding free energy of echinomycin to dsDNA as a function of ligand concentration [Echi].} 
Caption as in Table~\ref{tab: EcoRI 1}. Here $n=1$.}\label{tab: echi g}
\end{table}

\clearpage

\addcontentsline{toc}{subsection}{Tab. S10: Number of experiments performed at different concentrations of echinomycin with hairpin NSP}
\begin{table}
\begin{tabular}{lll}
\hline\hline
 [Echi] (nM) & pulled molecules (number of cycles per molecule) & total cycles $N$ \\
 \hline
 100  & 6 (131, 145, 761, 492, 230, 567) & 2326\\
 300  & 8 (504, 506, 528, 455, 46, 320, 219, 275) & 2853 \\
 1000 & 8 (779, 560, 547, 814, 666, 342, 377, 519) & 4604 \\
 3000 & 7 (310, 150, 268, 872, 241, 584, 533) & 2958\\
 \hline
\end{tabular}
\caption{{\bf Number of experiments performed at different concentrations of echinomycin with hairpin NSP.}}\label{tab: echins N}
\end{table}

\clearpage

\addcontentsline{toc}{subsection}{Tab. S11: Contributions to the binding free energy of echinomycin to dsDNA as a function of ligand concentration [Echi]}
\begin{table}
\begin{tabular}{ccccccc}
\hline\hline
 $i$ &  $f_{B_i}$ (pN) & $f_U$ (pN) & $\DG_{n_iU}^0$  & $\DW_{n_iU}^{\rm handles}+\DW_{n_iU}^{\rm bead}$ & $\DW_{n_iU}^{\rm ssDNA}$ & $\DW_{n_iU}^d$ \\
 \hline
 0  & 10.73$\pm$0.06 & 10.17$\pm$0.05 & 19.9$\pm$0.2   & -26.5$\pm$0.4 & 8.20$\pm$0.04 & -1.345$\pm$0.006 \\
 1  & 10.69$\pm$0.06 & 10.17$\pm$0.05 & 16.8$\pm$0.2   & -24.3$\pm$0.3 & 7.59$\pm$0.04 & -1.340$\pm$0.006 \\
 2  & 10.64$\pm$0.06 & 10.17$\pm$0.05 & 13.7$\pm$0.2   & -22.1$\pm$0.3 & 6.90$\pm$0.03 & -1.336$\pm$0.006 \\
 3  & 10.60$\pm$0.06 & 10.17$\pm$0.05 & 11.64$\pm$0.07 & -20.0$\pm$0.3 & 6.39$\pm$0.03 & -1.332$\pm$0.006 \\
 4  & 10.55$\pm$0.06 & 10.17$\pm$0.05 & 10.5$\pm$0.1   & -17.8$\pm$0.3 & 5.79$\pm$0.03 & -1.327$\pm$0.006 \\
 5  & 10.50$\pm$0.06 & 10.17$\pm$0.05 &  8.43$\pm$0.02 & -15.6$\pm$0.4 & 5.19$\pm$0.02 & -1.323$\pm$0.006 \\
 6  & 10.48$\pm$0.06 & 10.17$\pm$0.05 &  4.63$\pm$0.06 & -13.8$\pm$0.5 & 4.61$\pm$0.02 & -1.320$\pm$0.006 \\
 7  & 10.43$\pm$0.06 & 10.17$\pm$0.05 &  2.5$\pm$0.1   & -11.6$\pm$0.5 & 4.02$\pm$0.02 & -1.316$\pm$0.006 \\
 8  & 10.38$\pm$0.06 & 10.17$\pm$0.05 &  1.4$\pm$0.1   &  -9.4$\pm$0.5 & 3.44$\pm$0.02 & -1.311$\pm$0.006 \\
 9  & 10.33$\pm$0.06 & 10.17$\pm$0.05 & -0.11$\pm$0.06 &  -7.2$\pm$0.6 & 2.87$\pm$0.01 & -1.307$\pm$0.006 \\
 10 & 10.29$\pm$0.06 & 10.17$\pm$0.05 & -1.18$\pm$0.03 &  -5.1$\pm$0.6 & 2.29$\pm$0.01 & -1.302$\pm$0.006 \\
 11 & 10.24$\pm$0.06 & 10.17$\pm$0.05 & -2.71$\pm$0.03 &  -2.9$\pm$0.7 & 1.73$\pm$0.01 & -1.296$\pm$0.006 \\
\end{tabular}
\begin{tabular}{cccccc}
\hline\hline
 [Echi] (nM) & $\phi^{B\to U}$ & $\phi^{U\to B}$ & $\log \left(\phi^{B\to U}/\phi^{U\to B}\right)$ & $\DG_{BU}$ & $\DGb$ \\
 \hline
100  & 0.70$\pm$0.07 & 0.12$\pm$0.03 & 1.9$\pm$0.2  & -4.5$\pm$0.2 & -2$\pm$1 \\
300  & 0.53$\pm$0.07 & 0.30$\pm$0.01 & 0.5$\pm$0.1  & -3.3$\pm$0.2 & -1$\pm$1 \\
1000 & 0.37$\pm$0.02 & 0.46$\pm$0.03 & -0.2$\pm$0.1 & -2.6$\pm$0.2 &  0$\pm$1 \\
3000 & 0.21$\pm$0.04 & 0.70$\pm$0.05 & -1.3$\pm$0.3 & -1.2$\pm$0.4 &  1$\pm$1 \\
 \hline
\end{tabular}
\caption{{\bf Contributions to the binding free energy of echinomycin to dsDNA as a function of ligand concentration [Echi].} 
The term $\N$ is taken equal to $\N=12$, which is equal to the number of basepairs of H.NSP. The term $\log\sum_i\exp\left(-\beta\Delta G_{n_iU}\right)$ is equal to 7.6$\pm$0.5~$\kBT$.}\label{tab: echins g}
\end{table}

\clearpage

\addcontentsline{toc}{subsection}{Tab. S12: Number of experiments performed at 3000~nM [Echi] with hairpins C and NC}
\begin{table}
\begin{tabular}{lll}
\hline\hline
 Hairpin & pulled molecules (number of cycles per molecule) & total cycles $N$\\
 \hline
 HC  & 7 (40, 348, 312, 324, 603, 352, 374) & 2353\\
 HNC & 6 (997, 300, 193, 325, 845, 125) & 2785\\
 \hline
\end{tabular}
\caption{{\bf Number of experiments performed at 3000~nM [Echi] with hairpins C and NC.}}\label{tab: echi c nc N}
\end{table}

\clearpage

\addcontentsline{toc}{subsection}{Tab. S13: Contributions to the binding free energy for double binding to dsDNA at 3000 nM echinomycin1}
\begin{table}
\begin{tabular}{lcccccc}
\hline\hline
 Hairpin & $f_{B^2}$ (pN) & $f_U$ (pN) & $\DG_{{n_2}U}^0$  & $\DW_{{n_2}U}^{\rm handles}+\DW_{n_2U}^{\rm bead}$ & $\DW_{n_2U}^{\rm ssDNA}$ & $\DW_{B^2U}^d$ \\
 \hline
 C  & 20.0$\pm$0.2   & 19.4$\pm$0.2   & 18.3$\pm$0.8 & -51.1$\pm$0.4 & 12.10$\pm$0.07 & -1.967$\pm$0.008 \\
 NC & 16.67$\pm$0.05 & 15.94$\pm$0.05 & 18.3$\pm$0.8 & -41$\pm$1     & 10.56$\pm$0.03 & -1.785$\pm$0.003 \\
 \hline\hline
 \end{tabular}
 \begin{tabular}{cccccc}
 Hairpin & $\phi^{B^2\to U}$ & $\phi^{U\to B^2}$ & $\log \left(\phi^{B^2\to U}/\phi^{U\to B^2}\right)$ & $\DG_{B^2U}$ & $\DGb$ \\
 \hline
 C  & 0.26$\pm$0.05 & 0.72$\pm$0.03 & -1.1$\pm$0.3 & -14$\pm$1 & 4.5$\pm$0.5 \\
 NC & 0.4$\pm$0.1   & 0.15$\pm$0.02 & 0.9$\pm$0.3  &  -8$\pm$1 & 2.6$\pm$0.4 \\
 \hline
\end{tabular}
\caption{\textbf{Contributions to the binding free energy for double binding to dsDNA at 3000 nM echinomycin.} At 3000 nM [Echi] double binding events of echinomycin to hairpins C and NC are always observed. $n_2=4$ is the number of open basepairs when two echinomycin molecules bind to specific sites in H.C or H.NC.}\label{tab: echi c nc g}
\end{table}

\clearpage

\addcontentsline{toc}{subsection}{Tab. S14: Contributions to the binding free energy for single binding to dsDNA at 3000 nM echinomycin}
\begin{table}
\begin{tabular}{lcccccc}
\hline\hline
 Hairpin & $f_B$ (pN) & $f_U$ (pN) & $\DG_{n_1U}^0$  & $\DW_{n_1U}^{\rm handles}+\DW_{n_1U}^{\rm bead}$ & $\DW_{n_1U}^{\rm ssDNA}$ & $\DW_{n_1U}^d$ \\
 \hline
 NC & 16.1$\pm$0.2   & 15.95$\pm$0.06 & 6.6$\pm$0.4 & -15$\pm$2      & 5.65$\pm$0.01  & -1.757$\pm$0.006 \\
\hline\hline
 \end{tabular}
 \begin{tabular}{ccccc}
 $\phi^{B\to U}$ & $\phi^{U\to B}$ & $\log \left(\phi^{B\to U}/\phi^{U\to B}\right)$ & $\DG_{BU}$ & $\DGb$ \\
 \hline
 0.6$\pm$0.2   & 0.031$\pm$0.006 & 3.1$\pm$0.3 & -9$\pm$1 & -5$\pm$2\\
 \hline
\end{tabular}
\caption{\textbf{Contributions to the binding free energy for single binding to dsDNA at 3000 nM echinomycin.} Even though when pulling hairpin NC at 3000 nM [Echi] all binding events correspond to double binding, single binding events are also transiently observed with hairpin NC at large forces when the echimoycin molecule bound at the start of the hairpin stem spontaneously unbinds. Forward and reverse trajectories connecting such transient single binding state and the unfolded state also provide a measurement of $\DGb$ for hairpin H.NC.
} \label{tab: EchiS-SNC}
\end{table}

\clearpage

\addcontentsline{toc}{subsection}{Tab. S15: Number of experiments performed at 10~$\mu$M [Echi] with hairpins M}
\begin{table}
\begin{tabular}{ll}
\hline\hline
 pulled molecules (number of cycles per molecule) & total cycles $N$\\
 \hline
 3 (101, 300, 314) & 715\\
 \hline
\end{tabular}
\caption{{\bf Number of experiments performed at 10~$\mu$M [Echi] with hairpins M.}}\label{tab: echi m N}
\end{table}

\clearpage

\addcontentsline{toc}{subsection}{Tab. S16: Double binding of two echinomycin molecules to hairpin M stabilizes a misfolded state}
\begin{table}
\begin{tabular}{cccccc}
\hline\hline
 $f_N$ (pN) & $f_M$ (pN) & $\DG_{NM}^0$  & $\DW_{NM}^{\rm handles}+\DW_{NM}^{\rm bead}$ & $\DW_{NM}^{\rm ssDNA}$ & $\DW_{NM}^d$ \\
 \hline
7.32$\pm$0.03 & 6.75$\pm$0.05 & 23$\pm$1 & -16$\pm$1 & 3.48$\pm$0.02 & 0.606$\pm$0.006 \\
\end{tabular}
\begin{tabular}{cccccc}
\hline\hline
$\phi^{N\to M}$ & $\phi^{N\to M}$ & $\log \left(\phi^{N\to M}/\phi^{M\to N}\right)$ & $\DG_{NM}$ & $\DG_{\text{bind,}NM}$ \\
 \hline
 0.46$\pm$0.06 & 0.56$\pm$0.03 & -0.2$\pm$0.2 & 17$\pm$1 & 2$\pm$1 \\
 \hline
\end{tabular}
\caption{\textbf{Double binding of two echinomycin molecules to hairpin M stabilizes a misfolded state.} In the simultaneous binding of two echinomycin molecules to the folded hairpin M two unrelated structures are observed: one corresponds to the native state whereas the other correspond to a misfolded structure (Fig.~S1 and \ref{misfolding}). The difference in binding energy of echinomycin to each of the two structures is $\DG_{\text{bind,}NM}=2\pm1~\kBT$ according to our results.
}\label{tab: echi m g}
\end{table}


\clearpage
\begin{thebibliography}{10}

\bibitem{stormo2010determining}
{{G.~D. Stormo, \& Y. Zhao.} Determining the specificity of protein--{DNA} interactions. 
\textit{Nat. Rev. Genet.} \textbf{11}}, {751--760} (2010).

\bibitem{leavitt2001direct}
{S. Leavitt, \& E. Freire.} Direct measurement of protein binding energetics by isothermal titration calorimetry.
\textit{Curr. Opin. Struct. Biol.} \textbf{11}, {560--566} (2001).

\bibitem{mcdonnell2001surface}
{J.~M. McDonnell.} Surface plasmon resonance: towards an understanding of the mechanisms of biological molecular recognition.
\textit{Curr. Opin. Chem. Biol.} \textbf{5}, {572--577} (2001).

\bibitem{kalodimos2004structure}
{C.~G. Kalodimos, \textit{et~al.}} Structure and flexibility adaptation in nonspecific and specific protein-{DNA} complexes.
\textit{Science} \textbf{305}, {386--389} (2004).

\bibitem{kim2013probing}
{S. Kim, \textit{et~al.}} Probing allostery through {DNA}. \textit{Science} \textbf{339}, {816--819} (2013).

\bibitem{bustamante2003ten} 
{C. Bustamante, Z. Bryant, \& S.B. Smith.} Ten years of tension: single-molecule DNA mechanics. \textit{Nature} \textbf{421}, {423--427} (2003).

\bibitem{junker2005influence}
{J.P. Junker, K. Hell, M. Schlierf, W. Neupert, \& M. Rief.} 
Influence of substrate binding on the mechanical stability of mouse dihydrofolate reductase. \textit{Biophys. J.} \textbf{89}, {L46--L48} (2005).

\bibitem{cao2007functional}
{Y. Cao, M.M. Balamurali, D. Sharma, \& H. Li.} A functional single-molecule binding assay via force spectroscopy. \textit{Proc. Natl. Acad. Sci. U.S.A.} \textbf{104}, {15677--15681} (2007).

\bibitem{koirala2011single}
{D. Koirala, \textit{et~al.}} A single-molecule platform for investigation of interactions between {G}-quadruplexes and small-molecule ligands.
\textit{Nature Chem.} \textbf{3}, {782--787} (2011).

\bibitem{ainavarapu2005ligand}
{S.~R.~K. Ainavarapu, L. Lewyn, C. Badilla, \& J.M. Fernandez} Ligand binding modulates the mechanical stability of dihydrofolate reductase
\textit{Biophys. J.} \textbf{89}, {3337--44} (2005).

\bibitem{hann2007effect}
H. Eleanore, \textit{et~al.} 
The effect of protein complexation on the mechanical stability of Im9.
\textit{Biophys. J.} \textbf{92}, {L79--L81} (2007).

\bibitem{jarzynski1997nonequilibrium}
{C. Jarzynski.} Nonequilibrium equality for free energy differences.
\textit{Phys. Rev. Lett.} \textbf{78}, {2690} (1997).

\bibitem{crooks2000path}
{G.~E. Crooks.} Path-ensemble averages in systems driven far from equilibrium.
\textit{Phys. Rev. E} \textbf{61}, {2361} (2000).

\bibitem{liphardt2002equilibrium}
{J. Liphardt, S. Dumont, S.~B. Smith, I. Tinoco, \& C. Bustamante.}
Equilibrium information from nonequilibrium measurements in an experimental test of {J}arzynski's equality.
\textit{Science} \textbf{296}, {1832--1835} (2002).

\bibitem{collin2005verification}
{D. Collin, \textit{et~al.}} Verification of the crooks fluctuation theorem and recovery of {RNA} folding free energies.
\textit{Nature} \textbf{437}, {231--234} (2005).

\bibitem{shank2010folding}
{E.~A. Shank, C. Cecconi, J.~W. Dill, S. Marqusee, \& C. Bustamante.}
The folding cooperativity of a protein is controlled by its chain topology.
\textit{Nature} \textbf{465}, {637--640} (2010).

\bibitem{maragakis2008differential}
{P. Maragakis, M. Spichty, \& M. Karplus.}
A differential fluctuation theorem.
\textit{J. Phys. Chem. B} \textbf{112}, {6168--6174} (2008).

\bibitem{junier2009recovery}
{I. Junier, A. Mossa, M. Manosas, \& F. Ritort.}
Recovery of free energy branches in single molecule experiments.
\textit{Phys. Rev. Lett.} \textbf{102}, {070602} (2009).

\bibitem{alemany2012experimental}
{A. Alemany, A. Mossa, I. Junier, \& F. Ritort.}
Experimental free-energy measurements of kinetic molecular states using fluctuation theorems.
\textit{Nature Phys.} \textbf{8},
  {688--694} (2012).

\bibitem{suppmaterials}
{Materials and methods are available as supplementary materials on Science Online.}

\bibitem{orte2008direct}
A. Orte, \textit{et~al}.
Direct characterization of amyloidogenic oligomers by single-molecule fluorescence.
\textit{Proc. Natl. Acad. Sci. U.S.A.} \textbf{105}, {14424--14429} (2008).

\bibitem{fierz2012stability}
{B. Fierz, S. Kilic, A.~R. Hieb, K. Luger, \& T.~W. Muir.}
Stability of nucleosomes containing homogenously  ubiquitylated {H2A} and {H2B} prepared using semisynthesis.
\textit{J. Am. Chem. Soc.} \textbf{134}, {19548--19551} (2012).

\bibitem{yu2012direct}
H. Yu, \textit{et~al}. 
Direct observation of multiple misfolding pathways in a single prion protein molecule.
\textit{Proc. Natl. Acad. Sci. U.S.A.} \textbf{109}, {5283--5288} (2012).

\bibitem{lesser1990energetic}
D.~R. Lesser, M.~R. Kurpiewski, \& L. Jen-Jacobson.
The energetic basis of specificity in the {EcoRI} endonuclease--{DNA} interaction.
\textit{Science} \textbf{250}, {776--786} (1990).

\bibitem{jen1997protein}
L. Jen-Jacobson.
Protein--{DNA} recognition complexes: Conservation of structure and binding energy in the transition state.
\textit{Biopolymers} \textbf{44}, {153--180} (1997).

\bibitem{terry1983thermodynamic}
B. Terry, W. Jack, R. Rubin, \& P. Modrich.
Thermodynamic parameters governing interaction of {EcoRI} endonuclease with specific and nonspecific {DNA} sequences.
\textit{J. Biol. Chem.} \textbf{258}, {9820--9825} (1983).

\bibitem{koch2002probing}
S.~J. Koch, A. Shundrovsky, B.~C. Jantzen, \& M.~D. Wang. 
Probing protein-{DNA} interactions by unzipping a single {DNA} double helix.
\textit{Biophys. J.} \textbf{83}, {1098--1105} (2002).

\bibitem{santalucia1998unified}
J. SantaLucia.
A unified view of polymer, dumbbell, and oligonucleotide DNA nearest-neighbor thermodynamics.
\textit{Proc. Natl. Acad. Sci. U.S.A.} \textbf{95}, {1460--1465} (1998).

\bibitem{huguet2010single}
J.~M. Huguet, \textit{et~al}.
Single-molecule derivation of salt dependent base-pair free energies in {DNA}.
\textit{Proc. Natl. Acad. Sci. U.S.A.} \textbf{107}, {15431--15436} (2010).

\bibitem{van1984echinomycin}
M.~M. Van~Dyke, \& P.~B. Dervan.
Echinomycin binding sites on {DNA}.
\textit{Science} \textbf{225}, {1122--1127} (1984).

\bibitem{bailly1996cooperativity}
C. Bailly, F. Hamy, \& M.~J. Waring.
Cooperativity in the binding of echinomycin to DNA fragments containing closely spaced {CpG} sites.
\textit{Biochemistry} \textbf{35}, {1150--1161} (1996).

\bibitem{leng2003energetics}
{F. Leng, J.~B. Chaires, \& M.~J. Waring.}
Energetics of echinomycin binding to {DNA}.
\textit{Nucleic Acids Res.} \textbf{31}, {6191--6197} (2003).

\bibitem{heidarsson2014direct}
{P.~O. Heidarsson, \textit{et~al}.}
Direct single-molecule observation of calcium-dependent misfolding in human neuronal calcium sensor-1.
\textit{Proc. Natl. Acad. Sci. U.S.A.} \textbf{111}, {13069--13074} (2014).

\bibitem{thaastrom2004histone}
{A. Th{\aa}str{\"o}m, J. Gottesfeld, K. Luger, \& J. Widom.}
Histone-{DNA} binding free energy cannot be measured in dilution-driven dissociation experiments.
\textit{Biochemistry} \textbf{43}, {736--741} (2004).

\bibitem{schnell2004reaction}
S. Schnell, \& T.~E. Turner. 
Reaction kinetics in intracellular environments with macromolecular crowding: simulations and rate laws.
\textit{Prog. Biophys. Mol. Bio.} \textbf{85}, {235--260} (2004).

 \bibitem{forns2011improving} 
 {N. Forns, \textit{et~al}.}
Improving signal/noise resolution in single-molecule experiments using molecular constructs with short handles. \textit{Biophys. J.} \textbf{100}, {1765--1774} (2011). 
 
 \bibitem{cheng2011single} 
W. Cheng, S.~G. Arunajadai, J.~R. Moffitt, I. Tinoco, \& C. Bustamante. 
Single base pair unwinding and asynchronous RNA release by the hepatitis C virus NS3 helicase. 
\textit{Science} \textbf{333}, {1746--1749} (2011).
 
\bibitem{sidorova1996differences} 
{N.~Y. Sidorova, \& D.~C. Rau.} 
Differences in water release for the binding of EcoRI to specific and nonspecific DNA sequences.  
\textit{Proc. Natl. Acad. Sci. U.S.A.} \textbf{93}, {12272--12277} (1996).

 \bibitem{mfold} 
 {M. Zuker.} 
 Mfold web server for nucleic acid folding and hybridization prediction.
 \textit{Nucleic Acids Res.} \textbf{31}, {3406--3415} (2003).
 
 \bibitem{bennett} 
 C.~H. Bennett. 
 Efficient estimation of free energy differences from Monte Carlo data. 
 \textit{J. Comput. Phys.} \textbf{22}, {245--268} (1976).
 
 \bibitem{shirts} 
 M.~R. Shirts, E. Bair, G. Hooker, \& V.~S. Pande. 
 Equilibrium free energies from nonequilibrium measurements using maximum likelihood methods. 
 \textit{Phys. Rev. Lett.} \textbf{91}, 140601 (2003).
%
\bibitem{comstock2011ultrahigh}
{M.~J. Comstock, T. Ha, \& Y.~R. Chemla.}
Ultrahigh-resolution optical trap with single-fluorophore sensitivity.
\textit{Nat. Methods} \textbf{8}, {335--340} (2011).
%
 \bibitem{joanthio} J. Camunas-Soler, \textit{et~al}.
Single-molecule kinetics and footprinting of DNA bis-intercalation: the paradigmatic case of Thiocoraline.  \textit{Nucleic Acids Res.} \textbf{43}, {2767--2779} (2015).
%
\bibitem{alemany14} A. Alemany, \& F. Ritort. 
Determination of the elastic properties of short ssDNA molecules by mechanically folding and unfolding DNA hairpins. \textit{Biopolymers} \textbf{101}, 1193--1199 (2014).


\end{thebibliography}
\end{document}